\documentclass[iop,preprint,10pt]{emulateapj}
\usepackage[]{natbib}
\usepackage{graphicx}
\usepackage{subfigure}
\usepackage[dvipsnames]{color}
\usepackage{float}
\usepackage{amsmath}
\usepackage{fancyref}
\definecolor{orange}{RGB}{255,69,0}
\definecolor{green}{RGB}{0,255,0}
\definecolor{darkred}{RGB}{139,0,0}
\usepackage{amssymb}
\usepackage{lineno}
\usepackage[colorlinks=true,
            linkcolor=orange,
            urlcolor=magenta,
            citecolor=blue]{hyperref}
\usepackage{nameref}
\usepackage{array,multirow,makecell}
\usepackage{tabularx}

\usepackage{morefloats}
\usepackage{gensymb}

\begin{document}
\title
{\textit{Fermi}-Large Area Telescope observations of the brightest Gamma-ray flare ever detected from CTA 102 }

\author{Raj Prince$^{1}$, Gayathri Raman$^{1}$, Joachim Hahn$^{2}$, Nayantara Gupta$^{1}$, Pratik Majumdar$^{3}$ }
\affil{$^{1}$Raman Research Institute, Sadashivanagar, Bangalore 560080, India \\
        $^{2}$Max-Planck-Institut f\"ur Kernphysik, P.O. Box 103980, 69029 Heidelberg, Germany\\
        $^{3}$Saha Institute of Nuclear Physics, HBNI, Kolkata, West Bengal 700064, India} 

\email{rajprince@rri.res.in}
\begin{abstract}
We present a multi-wavelength study of the FSRQ CTA 102 using \textit{Fermi}-LAT and simultaneous Swift-XRT/UVOT observations. 
The \textit{Fermi}-LAT telescope detected one of the brightest flares from this object during Sep, 2016 to Mar, 2017.
In the 190 days of observation period the source underwent four major flares. A detailed analysis of the temporal and 
spectral properties of these flares indicates  the flare at MJD 57751.594 has a $\gamma$-ray   
flux of (30.12$\pm$4.48)$\times 10^{-6}$ ph cm$^{-2}$ s$^{-1}$ (from 90 minutes binning) in the energy range of 0.1--300 GeV.
This has been found to be the highest flux ever detected from CTA 102. 
Time dependent leptonic modelling of the pre-flare, rising state, flares and decaying state has been done. A single emission region of size $6.5\times 10^{16}$ cm
has been used in our work to explain the multi-wavelength spectral energy distributions. During flares the luminosity in electrons increases nearly 
seventy times compared to the pre-flare state.

\end{abstract}

\section{Introduction}
The blazar CTA 102 (FSRQ) is a luminous and well studied quasar located at a redshift of z = 1.037 (\citealt{Schmidt (1965)}).
Like other blazars it also has a jet, oriented close to our line of sight. Because of the relativistic beaming effect all the 
emission is beamed along the jet axis and as a result we observe a violent variability at all wavelengths.
As a quasar it was first identified by \citet{Sandage et al. (1965)} and classified as a highly polarised quasar by 
\citet{Moore et al. (1981)}. It is highly variable in the optical band and its variability has been investigated by 
\citet{Osterman Meyer et al.(2009)}, who found that faster variability is associated with higher flux states. 
Due to this behaviour, the object has been termed as an Optically Violent Variable (OVV) quasar (\citealt{Maraschi et al. (1986)}).
Variability in the form of flares has also been observed for this source in centimeter and millimeter wavelengths as well as in 
the X-rays (RXTE observation, \citealt{Osterman Meyer et al.(2009)}).

CTA 102 has also been observed in the $\gamma$-ray energy band by CGRO/EGRET and \textit{Fermi}-LAT telescopes where the source luminosity was 
observed to be $L_{\gamma}$= $5\times10^{47}$ erg/s, hence listed as a $\gamma$-ray bright source (\citealt{Nolan et al. (1993)}; \citealt{Abdo et al. (2009)}).
The Very Large Array (VLA) observation revealed its kpc-scale radio morphology with a strong central core and  two less 
luminous lobes on opposite sides (\citealt{Spencer et al. (1989)}; \citealt{Stanghellini et al. (1998)}). 
The Very Long Baseline Array (VLBA) 2 cm survey and its successor, the MOJAVE (MOnitering of Jets in Active galactic nuclei 
with VLBA Experiments)
program have been regularly monitoring CTA 102 since mid 1995. 
 The radio flare from CTA 102 observed around 2006  was studied in the following papers \citet{Fromm et al. (2011)}, \citet{Fromm et al. (2013a)},
\citet{Fromm et al. (2013b)}). VLBA  data collected over a period of May 2005 to  April 2007 spanning a frequency range of  2 to 86 GHz was
used to study the physical and kinetic properties of the jet. 
The apparent speeds of the various regions along the jet were estimated to be in the range of $0.77\pm 0.14c$ to $13.0\pm2.1c$ 
(\citealt{Fromm et al. (2013a)}).
The authors concluded that the variation in physical properties during flare was connected to a new travelling feature and the interaction 
between the shock wave and a stationary structure (\citealt{Fromm et al. (2013b)}). 

Results from the MOJAVE observations of the source morphology and kinematics at 15 GHz 
suggest apparent speed of the jet between 1.39c and 8.64c (\citealt{Lister et al. (2013)}). VLBA observations at 43 GHz revealed even higher 
apparent speed  of 18c (\citealt{Jorstad et al. (2001), Jorstad et al. (2005)}).  The variability Doppler  factor  was estimated to be $22.3\pm4.5$
by \citet{Jorstad et al. (2005)}. Two quasi-stationary components previously observed by \citet{Jorstad et al. (2005)} and the five moving 
components in the jet observed by \citet{Casadio et al. (2015)} N1, N2, N3, N4 and S1 revealed  the details of the jet structure and its kinetic
properties. The apparent speeds of the moving components N1, N2, N3 and N4 were reported as $14.9\pm0.2c$,  $19.4\pm0.8c$, $26.9\pm1.8c$ and 
$11.3\pm1.2c$ by \citet{Casadio et al. (2015)}.
They estimated the corresponding variability Doppler factors of these components as 14.6, 22.4, 26.1 and  30.3 respectively.  \\
Recently \citet{Li et al. (2018)} have studied the variability of CTA 102 covering the 2016 January flare. New 15 GHz, archival 43 GHz VLBA data  and the variable optical
flux density, degree of polarisation, electric vector position angle (EVPA) have been used to infer the properties of its jet. They inferred 
the Lorentz bulk factor of the jet to be more than 17.5 and intrinsic half opening angle less than $1.8^{\circ}$  from VLBA  data. \\

A prominent flare of flux $F_{> 100 MeV}$ = $5.2\pm0.4 \times 10^{-6}$ ph cm$^{-2}$ s$^{-1}$ (\citealt{Casadio et al. (2015)})
was detected by the LAT in Sep-Oct 2012. An optical and near-infrared (NIR) outburst was also simultaneously detected.

They found that the $\gamma$-ray outburst was coincident with outbursts in all frequencies and related to the passage of a new superluminal knot through the radio core.
During the flare the optical polarised emission showed intra-day variability. 

The late 2016 activity of CTA 102 was associated with intra-night variability in optical fluxes (\citealt{Bachev et al. (2017)}).
During the late 2016 to early 2017 high state the brightest flare from this source has been observed.

In this work we have studied the high state between Sep 2016 to Mar 2017  using  $\gamma$-ray and X-ray/UVOT data 
to explore the fast variability and time dependent multi-frequency spectral energy distributions (SEDs).

Throughout the paper we adopt the $\gamma-ray$ flux in units of $10^{-6}$ ph cm$^{-2}$ s$^{-1}$ unless otherwise mentioned.
We have used the flat cosmology model with H$_{0}$ = 69.6 km s$^{-1}$ Mpc$^{-1}$, and $\Omega_{M}$ = 0.27 to estimate the luminosity distance
(d$_{L}$ = 7.08 $\times10^{9}$ pc).

\section{Observations and data reduction}

CTA 102 was observed using \textit{Fermi-LAT} and Swift-XRT/UVOT during Sep 2016 - Mar 2017, details of which are given in Table \ref{Table:T1}.

\begin{table}
 \centering
 \begin{tabular}{ccc p{1cm}}
 \hline
 &&& \\
 Observatory & Obs-ID & Date (MJD) & Exposure (ks)\\
 &&& \\
 \hline
 &&& \\
 \textit{Fermi-LAT} & & 57650-& \\
 &&-57840& \\
 Swift-XRT      & 00033509081 & 57651 & 1.0\\
 Swift-XRT      & 00033509082 & 57657 & 0.8\\
 Swift-XRT      & 00033509085 & 57675 & 1.0\\
 Swift-XRT      & 00033509086 & 57681 & 1.0\\
 Swift-XRT/UVOT & 00033509087 & 57688 & 1.9\\
 Swift-XRT/UVOT & 00033509088 & 57689 & 1.7\\
 Swift-XRT/UVOT & 00033509090 & 57690 & 1.7\\
 Swift-XRT/UVOT & 00033509091 & 57691 & 1.6\\
 Swift-XRT/UVOT & 00033509092 & 57692 & 2.2\\
 Swift-XRT/UVOT & 00033509093 & 57706 & 2.9\\
 Swift-XRT/UVOT & 00033509094 & 57708 & 2.7\\
 Swift-XRT/UVOT & 00033509095 & 57710 & 3.1\\
 Swift-XRT/UVOT & 00033509096 & 57712 & 2.4\\
 Swift-XRT/UVOT & 00033509097 & 57714 & 1.7\\
 Swift-XRT/UVOT & 00033509098 & 57715 & 2.9\\
 Swift-XRT/UVOT & 00033509099 & 57719 & 1.9\\
 Swift-XRT/UVOT & 00033509101 & 57723 & 1.3\\
 Swift-XRT/UVOT & 00033509103 & 57728 & 1.9\\
 Swift-XRT      & 00033509105 & 57735 & 2.6\\
 Swift-XRT/UVOT & 00033509106 & 57738 & 2.4\\
 Swift-UVOT     & 00033509107 & 57740 & 0.8\\
 Swift-UVOT     & 00033509108 & 57742 & 2.4\\
 Swift-XRT/UVOT & 00033509109 & 57745 & 2.0\\
 Swift-XRT/UVOT & 00033509110 & 57748 & 1.6\\
 Swift-XRT/UVOT & 00033509111 & 57751 & 1.8\\
 Swift-XRT/UVOT & 00033509112 & 57752 & 1.4\\
 Swift-XRT/UVOT & 00088026001 & 57753 & 2.0\\
 Swift-XRT/UVOT & 00033509114 & 57754 & 1.4\\
 Swift-XRT/UVOT & 00033509113 & 57755 & 1.5\\
 Swift-XRT/UVOT & 00033509115 & 57759 & 2.4\\
 Swift-XRT/UVOT & 00033509116 & 57761 & 2.4\\
 Swift-XRT/UVOT & 00033509117 & 57763 & 2.5\\
 Swift-UVOT     & 00033509118 & 57765 & 0.5\\
 Swift-XRT/UVOT & 00033509119 & 57768 & 1.0\\
 Swift-XRT/UVOT & 00033509120 & 57771 & 1.7\\

 && & \\
 \hline
 
 \end{tabular}
 \caption{Table shows the log of the observations used for this work.}
 \label{Table:T1}
\end{table}

\subsection{Fermi-LAT}
 \textit{Fermi}-LAT is a pair conversion $\gamma$-ray Telescope sensitive to photon energies between 20 MeV 
to higher than 500 GeV, with a field of view of about 2.4 sr (\citealt{Atwood et al. (2009)}). The LAT's field of view covers about
20\% of the sky at any time, and it scans the whole sky every three hours.
The instrument was launched by NASA in 2008 into a near earth orbit. CTA 102 was continuously monitored by \textit{Fermi}-LAT since 2008 August. 
The first flare of CTA 102 was observed in Sep-Oct 2012 with the highest flux of $5.2\pm0.4$ (\citealt{Casadio et al. (2015)}).
The second flare observed during Sep 2016--Mar 2017 is brighter than the first.
 We have analysed the \textit{Fermi}-LAT data from 19 Sep 2016 to 31 Mar 2017
(MJD 57650-57840) to study  this flare. The standard data reduction and analysis procedure\footnote{https://fermi.gsfc.nasa.gov/ssc/data/analysis /documentation/} 
has been followed. The \textit{gtlike/pyLikelihood} method is used to analyse the data that is the part of the latest version 
(v10r0p5) of \textit{Fermi Science Tools}. The present analysis has been carried out after rejecting the events having zenith angle $>$ 90$\degree$,
in order to reduce the contamination from the Earth's limb  $\gamma$-rays. The latest Instrument Response Function (IRF) ``P8R2\_SOURCE\_V6"
has been used in the analysis. The photons are extracted from a circular region of radius 10$\degree$, with the region of interest (ROI)
centered at the position of CTA 102. The third \textit{Fermi}-LAT catalog (3FGL; \citealt{Acero et al. (2015)}) has been used to 
include the sources lying within a radius of 10$\degree$. 
All the spectral and flux parameters of the sources lying within 10$\degree$ from the center of ROI are left free to vary during model fitting.
Model file also includes the sources within 10$\degree$ to 20$\degree$ from the center
of ROI. However, all their spectral and flux parameters are kept fixed to the 3FGL catalog values.
It also includes the latest isotropic background model, ``iso\_P8R2\_SOURCE\_V6\_v06", and the galactic diffuse emission
model, ``gll\_iem\_v06",  both of which are standard models available from the \textit{Fermi} Science Support 
Center \footnote{http://fermi.gsfc.nasa.gov/ssc/data/access/lat /BackgroundModels.html} (FSSC). 
A maximum likelihood (ML) test has been done to test the significance of $\gamma$-ray signal. The ML is defined as TS = 2$\Delta$log(L),
where L is the likelihood function between models with and without a point source at the position of the source of interest 
(\citealt{Paliya (2015)}).
The ML analysis was performed over the period of our interest and the 
sources which fall below 3$\sigma$ detection limit (i.e. TS $<$ 9; for details see \citealt{Mattox et al. (1996)}) are removed
from the model file.\\

\par
  The  spectral properties of the \textit{Fermi}-LAT detected blazars 
are most often studied by fitting their differential gamma ray spectrum with the following functional forms 
\citep{Abdo et al. (2010a)}. 
\begin{itemize}
\item A power law (PL), defined as
\begin{equation} \label{1}
 dN(E)/dE = N_0 (E/E_p)^{-\Gamma},
 \end{equation}

\end{itemize}
with pivot energy (energy at which error on differential flux is minimal) $E_p$ = 476.0 MeV from 1FGL \citep{Abdo et al. (2010b)} 
\begin{itemize}
 \item A log-parabola (LP), defined as
 \begin{equation} \label{2}
 dN(E)/dE = N_0 (E/E_0)^{-\alpha-\beta\ln(E/E_0)},
 \end{equation}

\end{itemize}
with pivot energy $E_0$ = 308.3 MeV from 3FGL \citep{Acero et al. (2015)} 
where $\alpha$ is the photon index at $E_0$, $\beta$ is the curvature index and 
``\rm{ln}" is the natural logarithm;
\begin{itemize}
 \item A power law with an exponential cut-off (PLEC), defined as
 \begin{equation} \label{3}
  dN(E)/dE = N_0 (E/E_p)^{-\Gamma} \exp(-E/E_c),
  \end{equation}

\end{itemize}
with pivot energy $E_p$ = 476.0 MeV from 1FGL \citep{Abdo et al. (2010b)} 
\begin{itemize}
 \item A broken power law (BPL), defined as
 \begin{equation} \label{4}
  dN(E)/dE = N_0 (E/E_{break})^{{-\Gamma}_{i}},
  \end{equation}
\end{itemize}
with $i$ = 1 if $E$ $<$ $E_{break}$ and $i$ = 2 if $E$ $>$ $E_{break}$.

\subsection{Swift-XRT and UVOT}
CTA 102 was observed by Swift-XRT/UVOT during the flaring period of Sep 2016-Jan 2017 
(there has been no observation during Feb and Mar 2017). Details of the observations are presented in Table \ref{Table:T1}. 
 
Cleaned event files were obtained using the task `\textit{xrtpipeline}' version 0.13.2.
Latest calibration files (CALDB version 20160609) and standard screening criteria 
were used for re-processing the raw data. Cleaned event files corresponding to the Photon Counting (PC) mode were considered. 
Circular regions of radius 25 arc seconds centered at the source and slightly away from the source were 
 chosen for the source and the background regions respectively while analysing the XRT data.

The X-ray spectra were extracted in \textit{xselect}.
The tool \textit{xrtmkarf} was used to produce the ancillary response file and \textit{grppha} was used to group the spectra to obtain a minimum
of 30 counts per bin. The grouped spectra were loaded in XSPEC for spectral fitting.  All the spectra were 
fitted using an absorbed broken power law model with the galactic absorption column density $n_H$ = 5.0$\times$10$^{20}$ 
cm$^{-2}$ \citep{Kalberla et al. (2005)}. 

The Swift Ultraviolet/Optical Telescope (UVOT, \citealt{Roming et al. (2005)}) also observed CTA 102 in all the six filters U, V, B, W1, M2 and W2. 
The source image was extracted from a region of 10 arc seconds centered at the source. The background region was chosen 
with a radius of 30 arc seconds away from the source from a nearby source free region . The `uvotsource'  task has been used to extract 
the source magnitudes and fluxes. Magnitudes are corrected for galactic extinction (\citealt{Schlafly et al. (2011)}) and converted into flux
using the zero points (\citealt{Breeveld et al. (2011)}) and conversion factors (\citealt{Larionov et al. (2016)}).

\section{Results}
Our results on the light curves, variability time, and spectral analysis during the Sep 2016-Mar 2017 
flaring period of CTA 102 are presented in this section.
\subsection{Light curves - Gamma ray}
\begin{figure*}[htbp]
\begin{center}
\includegraphics[scale=0.50]{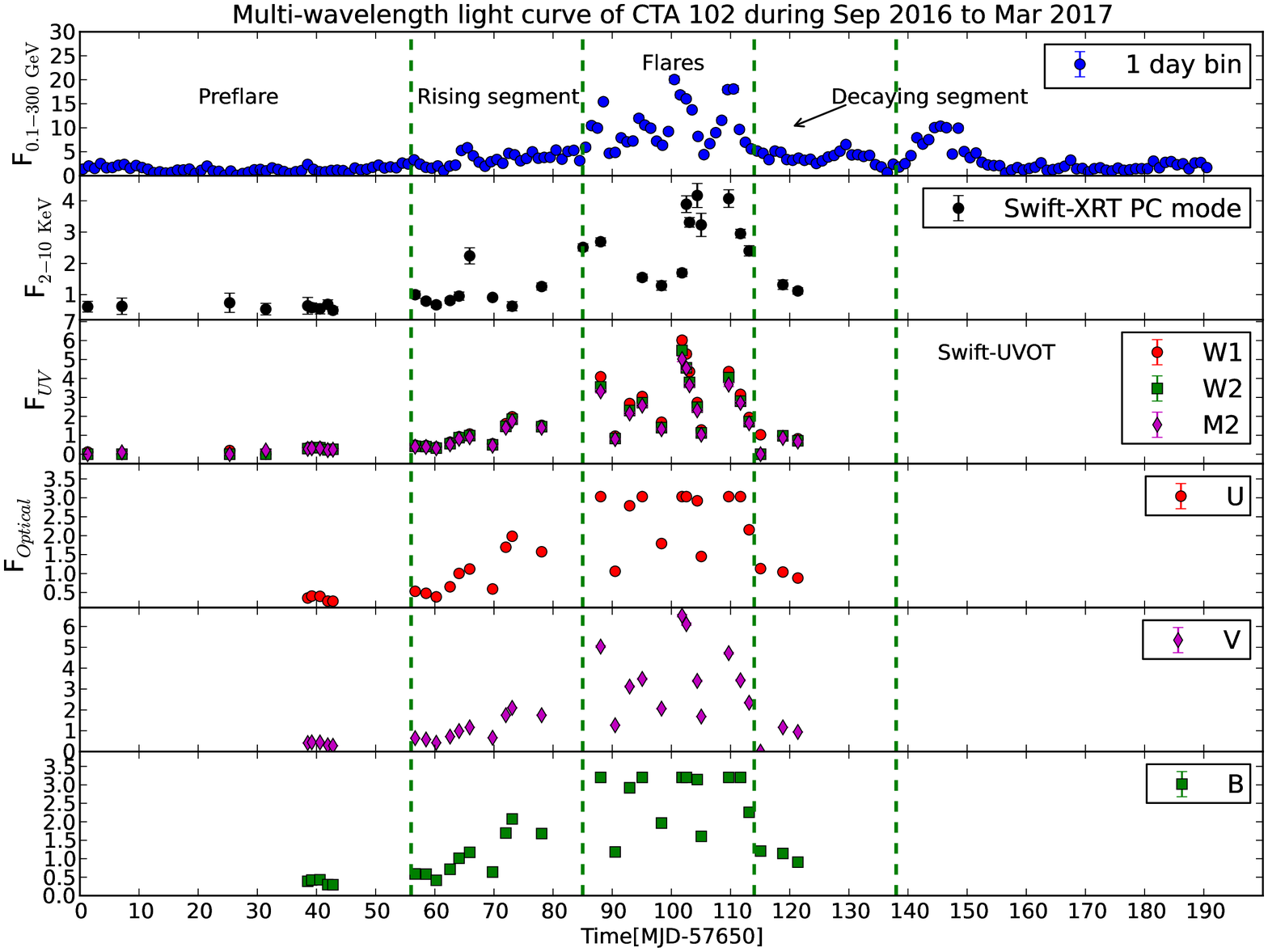}
\end{center}
\caption{Light curve of the CTA 102 during the Sep 2016--Mar 2017 outburst. Four major flaring episodes have been identified and 
further studied. Pre-flare, rising and decaying segments are also present before and after the flaring episodes, separated by green dashed line.
XRT is in unit of 10$^{-11}$, F$_{UV}$ and F$_{optical}$ is in unit of 10$^{-10}$ erg cm$^{-2}$ s$^{-1}$. $\gamma$-ray flux 
shown in the top panel is in unit of 10$^{-6}$ ph cm$^{-2}$ s$^{-1}$.}
\label{fig:A}

\begin{center}
\includegraphics[scale=0.5]{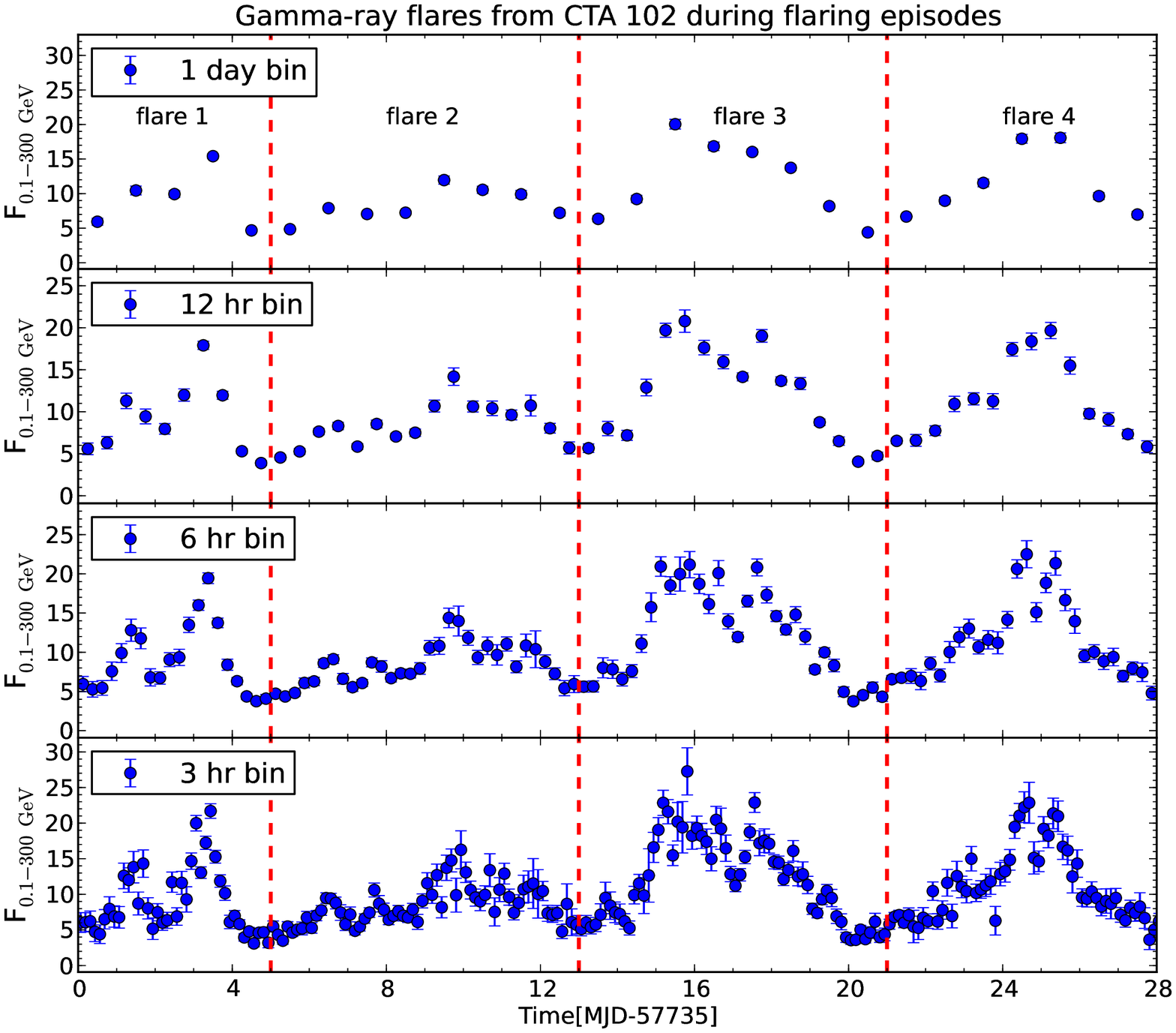}
\end{center}
\caption{Zoomed version of flaring episodes. The flares are separated by
red dashed lines and their time durations are as follows: MJD 57735--57740, MJD 57740--57748, MJD 57748--57756, and MJD 57756-57763. Sub-structures 
are clearly seen from 6 hr and 3 hr binning which also hints about the flux variability shorter than day scale. }
\label{fig:B}
\end{figure*}

The variability of the source can be  determined from its light curve. The light curve of CTA 102, observed by the 
$\textit{Fermi}$-LAT during Sep 2016-Mar 2017, has been shown in the top panel of Figure \ref{fig:A}. 
The $\gamma$-ray variability of the source can be clearly seen by generating the light curves (see Figure \ref{fig:B}) with various time bins 
(1 day, 12 hr, 6 hr, and 3 hr). The spectral analysis has been carried out, using the \textit{unbinned likelihood analysis} method,
over several periods of flaring states in the energy range of 0.1--300 GeV.
The whole light curve is divided into four parts: pre-flare, rising segment, flare and decaying segment separated by green dotted lines.
The source started showing high activity during MJD 57735 and went on till MJD 57763 (Figure \ref{fig:B}).
The flaring state which lasted for 28 days is divided into four flares.  
These four flares separated by  dotted red lines are shown in Figure \ref{fig:B}. We have named them as flare-1, 
flare-2, flare-3 and flare-4 which corresponds to MJD 57735--57740, MJD 57740--57748, MJD 57748--57756 and MJD 57756--57763
respectively. 
The quiescent states before and after the flaring period are identified as pre-flare phase, rising segment and decaying segment respectively. 
The various time bins (1 day, 12 hr, 6 hr and 3 hr) are applied to study the behaviour of CTA 102 during the flaring period. 
In Figure \ref{fig:B}, the light curves for different time bins are shown in different panels. The top panel represents 1 day time binning, 
which reveals the four flares. The 12 hr, 6 hr and 3 hr binning reveals the substructures in each flare. These substructures are nothing but 
the collection of peaks of different heights. 
The data points shown in the bottom most panel (i.e. 3 hr binning) of Figure \ref{fig:B} are used to study
the variability time scale and the data points in the panel just above it (i.e. 6 hr binning) are used to study the temporal behaviour. The data points 
below the detection limit of 3$\sigma$ (TS $<$ 9; \citealt{Mattox et al. (1996)}) have been rejected from both the temporal and variability
study.
\par
The temporal evolution of each flare has been studied separately. For this purpose we have fitted the peaks, found in the 6 hr binning of light 
curve, by a sum of exponentials which gives the rise and decay times of each peak. The functional form of the sum of exponentials is as follows:
\begin{equation} \label{6}
 F(t) = 2F_0 \left[\rm{exp}(\frac{t_0-t}{T_r}) + exp(\frac{t-t_0}{T_d})\right]^{-1},
\end{equation}
 where $F_0$ is the flux at time $t_0$ representing the approximate flare amplitude, $T_r$ 
and $T_d$ are the rise and decay times of the flare respectively \citep{Abdo et al. (2010c)}.

Any physical process faster than the light travel time
or the duration of the event will not be detectable from the light curve (\citealt{Chiaberge and Ghisellini (1999)}; \citealt{Chatterjee et al. (2012)}).
A symmetric temporal evolution, having equal rise and decay times, may occur when a perturbation in the jet flow or a blob of denser plasma  
passes through a standing shock present in the jet (\citealt{Blandford and Konigl (1979)}).   
In flare-1 the first peak (P1) has nearly equal rise and decay times. The second peak (P2) has a longer rise time than decay time. This could 
be due to slow injection of electrons into the emission region. This is also seen in first peak (P1) and fourth peak (P4) of flare-2.
A slower decay time could be due to longer cooling time of electrons. In flare-2 the second peak (P2) and in flare-3 the fourth peak (P4) have
significantly longer decay time than rise time.
Among the 14 peaks of four flares given in Table \ref{Table:T2} five peaks have nearly equal rise and decay times. Five peaks have slower rise time than decay
time and the rest of the four have slower decay time than rise time. Hence all the three scenarios are almost equally probable.

\subsubsection{Pre-flare, Rising and decaying Segment}
Prior to its flaring activity in Dec 2016, CTA 102 was in quiescent state.
We call it as pre-flare phase. We have defined the pre-flare 
during MJD 57650 -- MJD 57706 (Figure \ref{fig:A}). The average flux during pre-flare phase is observed to be 
F$\rm_{GeV}$ = 1.40$\pm$0.10 ph/cm$^{2}$/s. Just after
the MJD 57706 the source flux started rising but very slowly. We call it as the rising segment with average flux  F$\rm_{GeV}$ = 3.83$\pm$0.08 
ph/cm$^{2}$/s, lasting for a period of one month MJD 57706 -- MJD 57735.
The flaring phase started from MJD 57735 and consisted of four major flares which lasted up to 28 days
(till MJD 57763) after which the emission started decaying very slowly. We name it as the decaying segment. The average flux in this 
period was almost similar to the rising segment.
In the following sections, we discuss the four flares in detail. 

\subsubsection{Flare-1}
Flare-1, as shown in Figure \ref{fig:B}, was observed during MJD 55735--57740 before which 
the source was in rising state as mentioned in Section 3.1.1 . 
The temporal evolution of flare-1 is shown in the left panel of Figure \ref{fig:D}. The 6 hr bins clearly reveal that the flux started rising 
just from MJD 55735 and lasted up to five days (till MJD 57740). There are two major peaks P1 and P2, during flare-1, at 
MJD 57736.375 and MJD 57738.375 with a flux of F$\rm_{GeV}$ = 12.81$\pm$1.42 and 19.45$\pm$0.68 respectively. These peaks are 
fitted with the function given in Equation \ref{6}. The rise and decay times of the peaks are found from this fit. 
The details of the fitted parameters are presented in Table \ref{Table:T2}. 
This has been done for all the four flares.
Along with the peaks we have also fitted the baseline flux, 
shown in Figure \ref{fig:D} (grey line), which is very close to the quiescent state mentioned in Section 3.1.1.
 
\subsubsection{Flare-2}
\begin{figure*}[htbp]
\begin{center}
\includegraphics[scale=0.53]{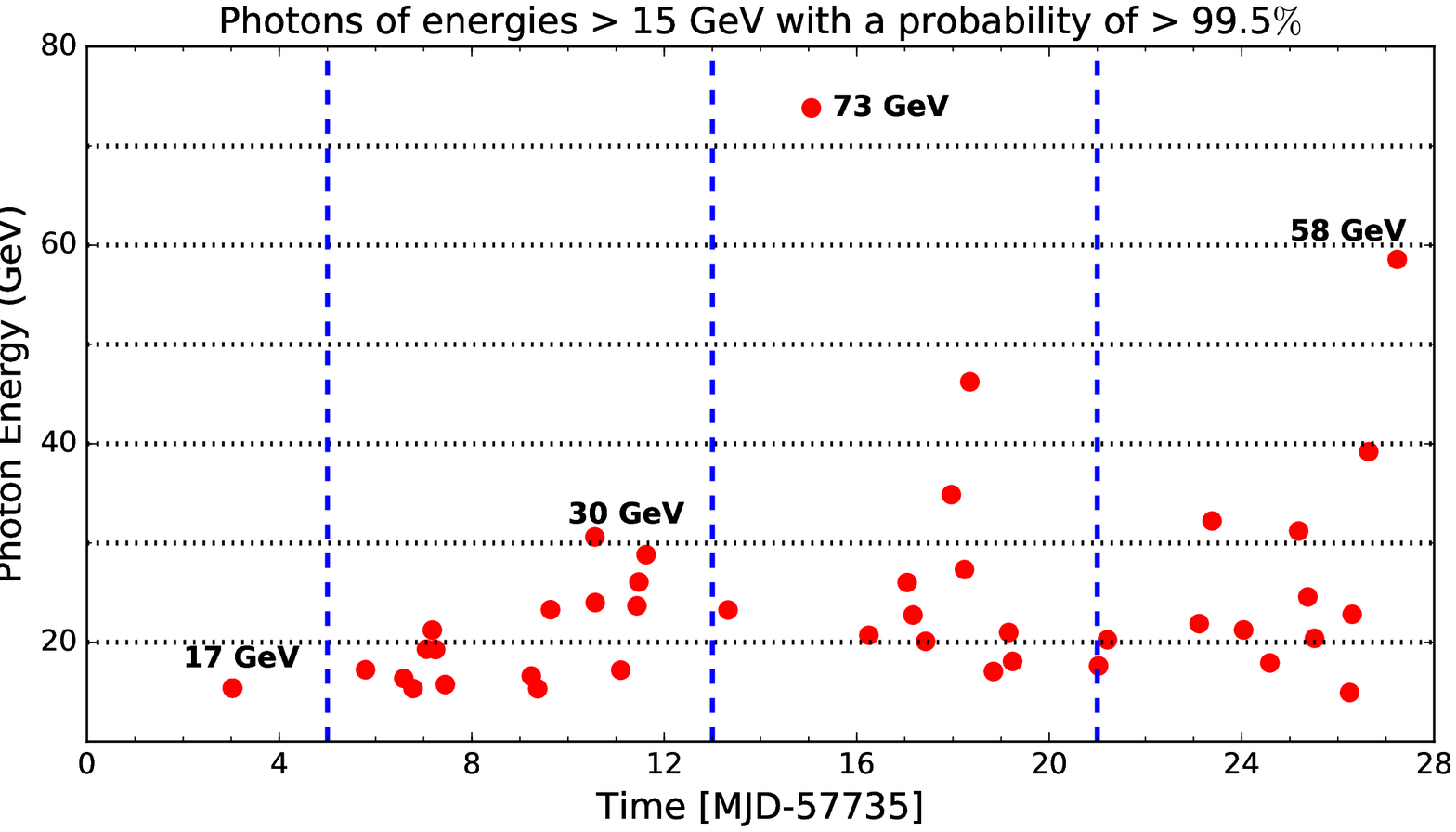}
\end{center}
\caption{Arrival time and energy of E $>$ 15 GeV photons, with probability greater than 99.5$\%$, plotted for all the flares. Vertical
dashed blue lines are separating the flares.} 

\label{fig:C}

\begin{center}
\includegraphics[scale=0.29]{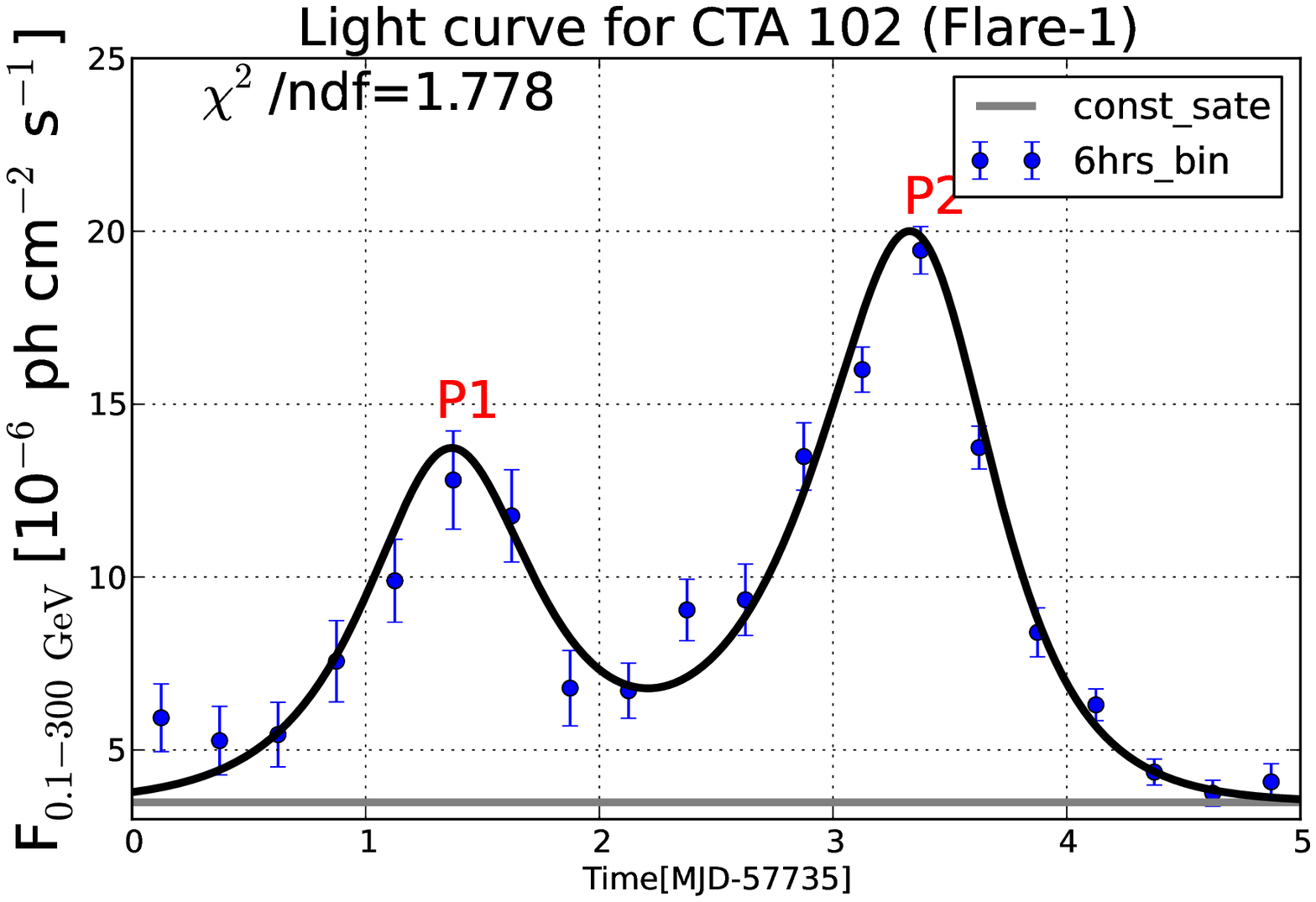}
\includegraphics[scale=0.29]{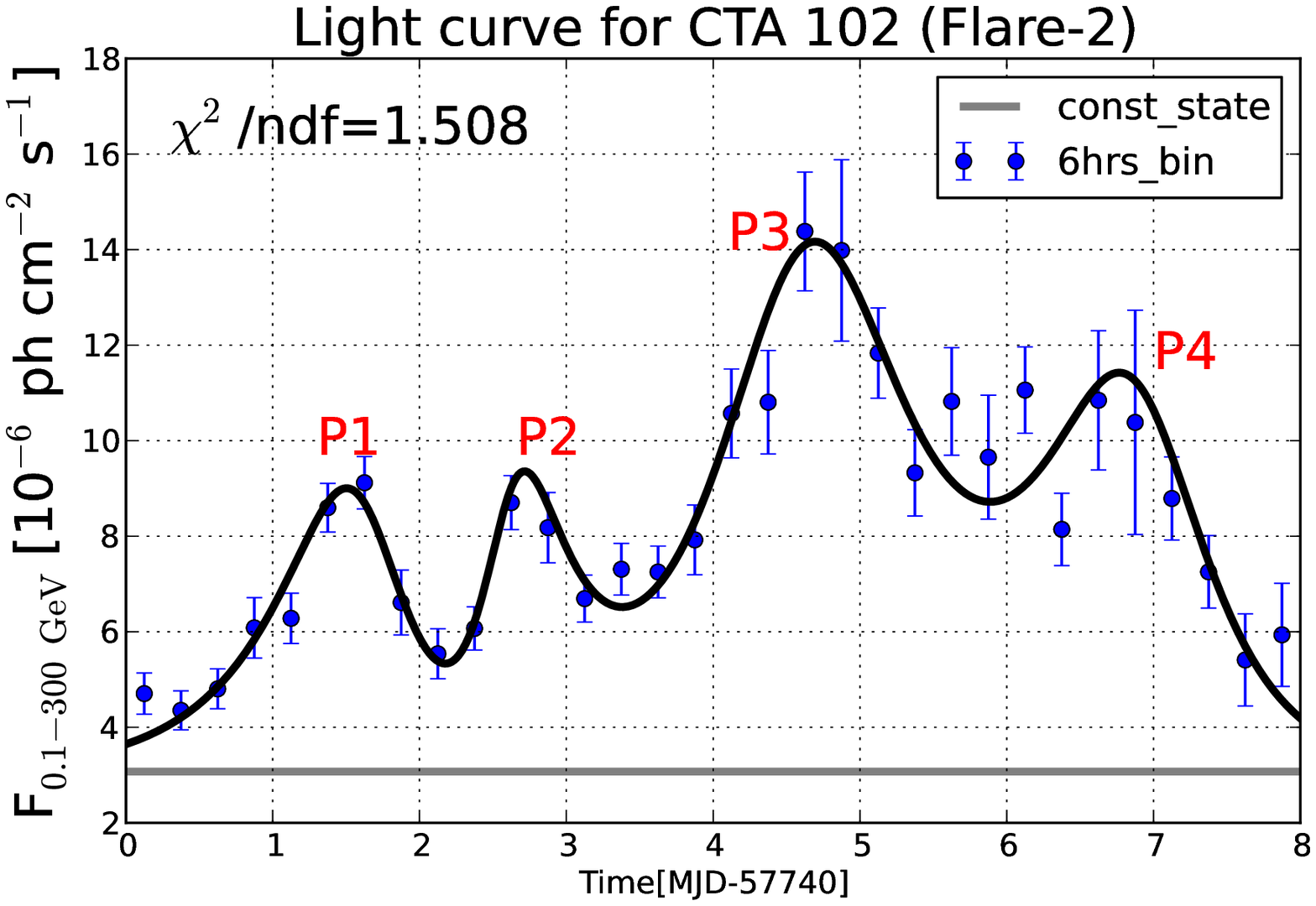}
\end{center}
\caption{left panel: Light curve of flare-1. right panel: Light curve of flare-2. The peaks of both the flares are fitted by 
the sum of exponentials and the fitted parameters are mentioned in Table \ref{Table:T2}. The light grey line represents the constant 
flux or baseline.}
\label{fig:D}

\begin{center}
\includegraphics[scale=0.29]{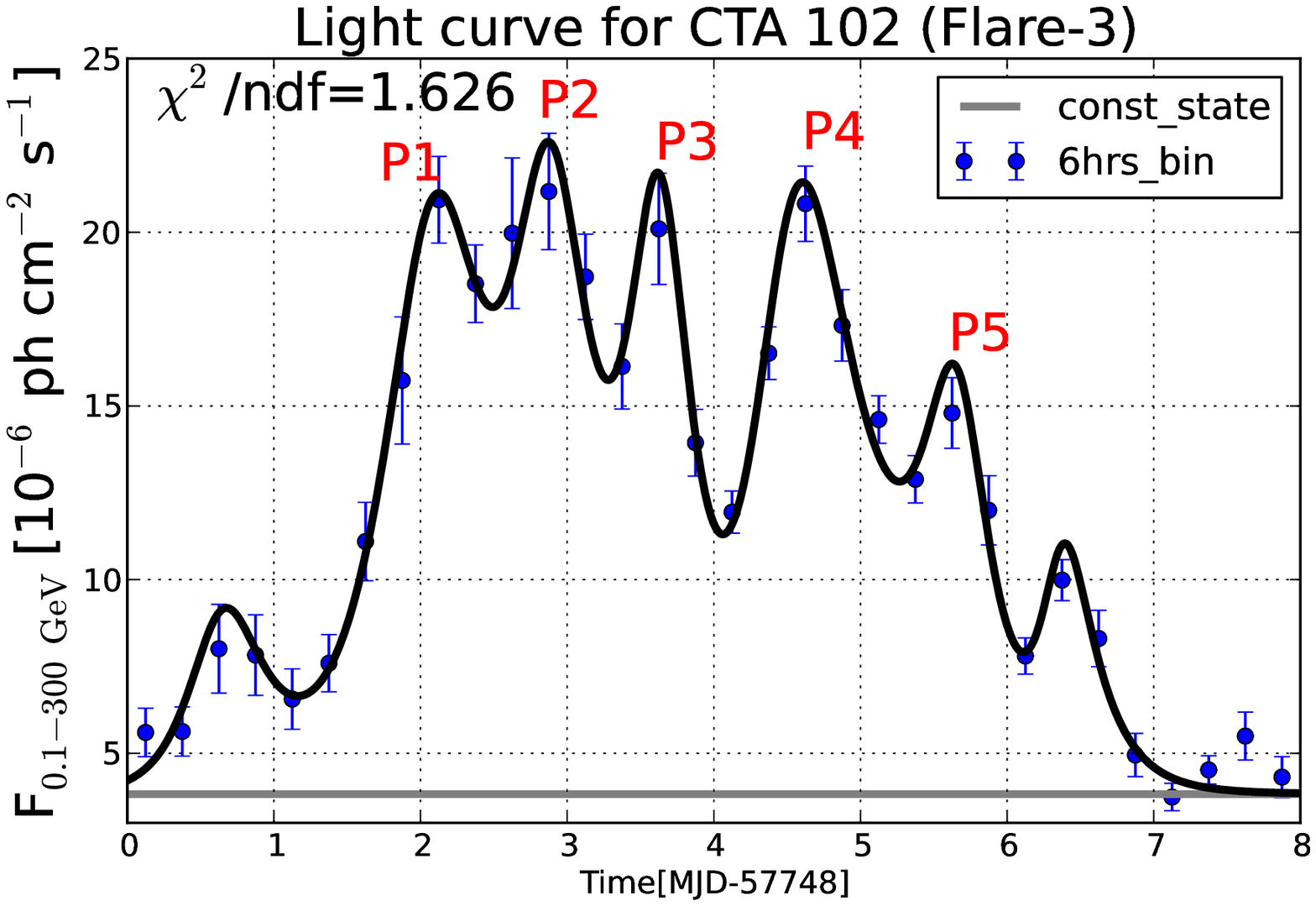}
\includegraphics[scale=0.29]{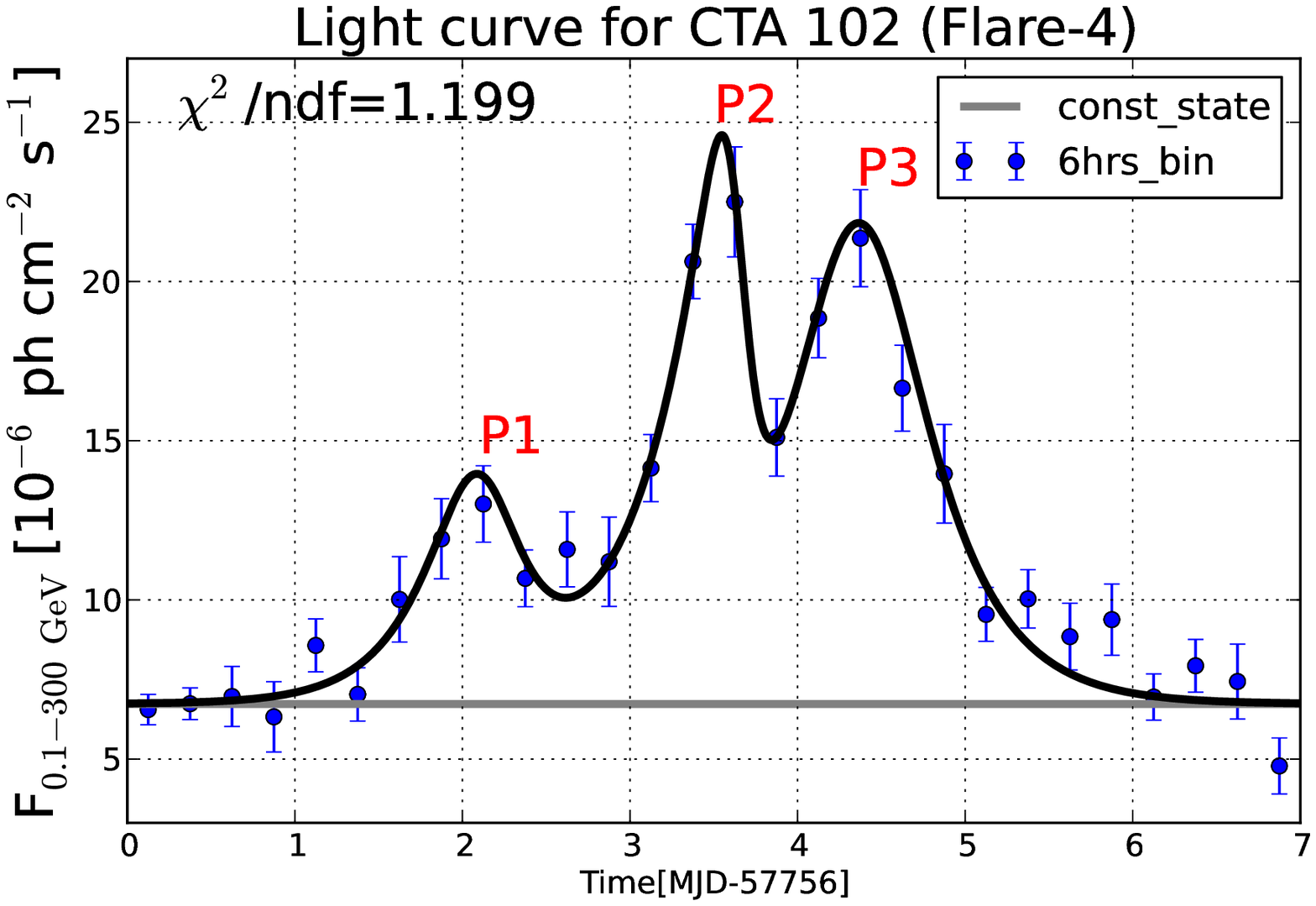}
\end{center}
\caption{left panel: Light curve of flare-3. right panel: Light curve of flare-4. The peaks of both the flares are fitted by 
the sum of exponentials and the fitted parameters are mentioned in Table \ref{Table:T2}. The light grey line represents the constant 
flux or baseline.}
\label{fig:E}
\end{figure*}

The temporal evolution of flare-2 (MJD 57740--57748) is shown in the right panel of Figure \ref{fig:D}. As flare-1 started to decay at 
MJD 55740, flare-2 started to rise and lasted up to eight days till MJD 57748. The flaring period of eight days can be clearly divided into
four major peaks (Figure \ref{fig:D}) called as P1, P2, P3 and P4 which happened at MJD 57741.625, 57742.625, 57744.625 and 57746.625 
with a flux of F$\rm_{GeV}$ = 9.12$\pm$0.55, 8.70$\pm$0.56, 14.38$\pm$1.24 and 10.84$\pm$1.46 respectively. The baseline flux shown in Figure \ref{fig:D} (grey line) is
close to the quiescent state flux.

\subsubsection{Flare-3}
Similar to flare-1 and flare-2 a 6 hr binning of flare-3 has been done to study the temporal evolution, as shown in left panel of Figure \ref{fig:E}.
Flare-3 was observed during MJD 57748--57756 and is one of the brightest flares ever detected from  
CTA 102 with a flux of F$\rm_{GeV}$ = 27.26$\pm$3.30 at MJD 57750.813 (from 3 hr binning). It is much brighter than the flare
observed in Sep-Oct 2012 (\citealt{Casadio et al. (2015)}) with a flux of F$\rm_{GeV}$ = 5.2$\pm$0.4.
The flux started to rise from the point where flare-2 diminished, i.e at MJD 57748. The source spent around
seven days in its flaring state (see Figure \ref{fig:E}) subsequently its flux  reduced to the quiescent
state flux value. The flaring period is divided into five major and clear peaks shown in left panel of Figure \ref{fig:E}. The peaks P1, P2, 
P3, P4 and P5 were observed at MJD 57750.125, 57750.875, 57751.625, 57752.625 and MJD 57753.625 with fluxes of 
F$\rm_{GeV}$ = 20.93$\pm$1.24, 21.17$\pm$1.67, 20.09$\pm$1.60, 20.82$\pm$1.08 and 14.79$\pm$1.01 respectively.

\subsubsection{Flare-4}
A 6 hr binning of flare-4 has also  been carried out during MJD 57756--57763 to study the temporal evolution. The light curve of flare-4
is shown in right panel of Figure \ref{fig:E}. The
flux started  to rise at MJD 57757
and stayed in the flaring state for around six days (MJD 57757--57763). Three major peaks were observed during the flaring period
of flare-4. We name them as P1, P2, and P3 which happened at MJD 57758.125, 57759.625, 57760.375 with the fluxes of F$\rm_{GeV}$ =
13.01$\pm$1.20, 22.50$\pm$1.73, and 21.36$\pm$1.52 respectively. 

\subsection{X-ray and UV/Optical light curves}

The simultaneous X-ray (2 -- 10 keV) and UVOT (six filters) light curves are shown in lower panel of Figure \ref{fig:A}. 
Swift observation for flaring episode was carried out along with \textit{Fermi}-LAT till MJD 57771 (Mar 18 2017).  
The 2-10 keV X-ray 
fluxes, obtained using the CFLUX convolution model, have been used to plot the X-ray light curve in Figure \ref{fig:A}. The UVOT light curves 
show the fluxes in all the six filter bands during each of the observations. The light curves, though sparsely populated as 
compared to the $\gamma$-ray light curves, do show correlated increased intensities during the flaring episodes.

\subsection{Variability }
The variability time is a measure of how fast the flux is changing with time during the flaring period.
It also provides an estimate of the size of the emission region for a given value of Doppler factor ($\delta$) of the jet and 
redshift (z) of the source. To estimate the fastest variability time we have done 90 minutes time binning and used the following function:
\begin{equation} \label{7}
F(t_2) = F(t_1).2^{(t_2-t_1)/ t_{d}},\\
\end{equation}
where $F(t_1)$ and $F(t_2)$ are the fluxes measured at two instants of time $t_1$ and $t_2$ respectively and $t_d$ represents 
the doubling/halving timescale of flux. We have scanned all the four flares shown in Figure \ref{fig:B} with the function 
given in Equation \ref{7}. While scanning the light curves we use the following conditions: Only those consecutive time instants will be
considered which have at least 5$\sigma$ detection (TS $>$ 25) and the flux between these two time instants should be 
double (rising part) or half (decaying part). There are time instants which have 5$\sigma$ detection but the 
difference in values of fluxes measured at these instants is less than a factor of two or vice-versa. These time instants are completely 
ignored in our fastest variability analysis. The shortest variability time is found to be t$_{var}$ = 1.08$\pm$0.01 hr between MJD 
57761.47 -- 57761.53; which is consistent with the hour scale variability found for other FSRQ like PKS 1510-089 (\citealt{Prince et al. (2017)}).
 We also found that the variability time during flares in \textit{Fermi}-LAT data varies in the range of 1 hour to several days.

\subsection{Fractional Variability (F$_{var}$)}
Fractional variability is used to determine the variability amplitudes across the whole electromagnetic spectrum (\citealt{Vaughan et al. (2003)})
during simultaneous multi-wavelength observations of blazars. But here we have calculated with only the gamma ray data to identify the 
different activity states of the blazar. The fractional variability amplitude was first introduced by (\citealt{Edelson and Malkan (1987)};
\citealt{Edelson et al. (1990)}) and it can be estimated by using the relation given in \citet{Vaughan et al. (2003)},
\begin{equation} \label{8}
 F_{var} = \sqrt{\frac{S^2 - \sigma^2}{r^2}} \\
\end{equation}

\begin{equation}
 err(F_{var}) = \sqrt{  \Big(\sqrt{\frac{1}{2N}}. \frac{\sigma^2}{r^2F_{var}} \Big)^2 + \Big( \sqrt{\frac{\sigma^2}{N}}. \frac{1}{r} \Big)^2     } \\
\end{equation}

where, $\sigma^2_{XS}$ = S$^{2}$ -- $\sigma^2$, is called excess variance, S$^{2}$ is the sample variance, $\sigma^2$ is the mean square uncertainties 
of each observations and r is the sample mean. We have also estimated the normalised excess variance, $\sigma^2_{NXS}$ = $\sigma^2_{XS}$/r$^2$.

\begin{table}
\caption{}
\begin{tabular}{c c c c c}
 \hline
 \\
 Activity & $\sigma^2_{NXS}$ & err($\sigma^2_{NXS}$) & F$_{var}$ & err(F$_{var}$) \\
 \\
 \hline
 \\
 Preflare & 0.0626 & 0.0192 & 0.2502 & 0.0384 \\ 
 \\
 Rising segment & 0.1295 & 0.0147 & 0.3598 & 0.0205 \\
 \\
 Decaying segment & 0.1189 & 0.0129 & 0.3449 & 0.0188 \\
 \\
 Flare-(1-4)   & 0.1899 & 0.0076 & 0.4358 & 0.0087  \\
 \\
 \hline
\end{tabular}
\label{Table:TA}

\end{table}

The rise in the values of the fractional variability and excess variance from pre-flare to flare state and subsequent fall during decaying state
are shown in Table \ref{Table:TA}.
 The fractional variabilities in multi-wavelength data have been calculated by \citet{Kaur and Baliyan (2018)} for the same source. 
They have found larger fractional variability at higher energy e.g. 0.87 in $\gamma$-rays, 0.45 in X-rays, 0.082 in UVW2-band
and 0.059 in optical B-band. Similar results have also been reported by \citet{Patel et al. (2018)} for 1ES 1959+650. They have found fractional variability increases
with increasing energy.
The opposite scenario was observed by \citet{Bonning et al. (2009)} for 3C 454.3, where fractional variability decreases 
with increasing energy (IR, Optical, UV) due to the presence of steady thermal emission from the accretion disk.

\subsection{High Energy Photons}
In Figure \ref{fig:C} we have plotted the photons energy ($>$ 15 GeV), with respect to their arrival time on x-axis, for all the flares shown in 
Figure \ref{fig:B}. To get the high energy photons the \textit{Fermi}-analysis has been done with the ``ULTRACLEAN'' class of events and 0.5$\degree$
of ROI. The high energy photons that have energy E $>$ 15 GeV and also the probability above 99.5$\%$ are only presented in Figure \ref{fig:C}. We 
find that a photon of energy E = 73.8 GeV was detected at MJD 57750.06 with a probability of 99.99$\%$, and this is part of the 
brightest flare ever detected from CTA 102 i.e. flare-3. 
This is the highest energy photon ever detected from CTA 102. Figure \ref{fig:C} clearly reveals that
most of the high energy photons are detected during flare-2, 3, and 4. 
Photons of energy 17, 30 and 58 GeV with probability of 99.99$\%$ also have been detected during flare-1, 2 and 4 at MJD 
57740.79, 57745.56 and 57762.23 respectively. 
 Such high energy photons can be produced in external Compton scattering of the BLR, disk or dusty torus photons by the relativistic electrons
in the jet and also by synchrotron self Compton emission.

\subsection{Spectral Energy distributions of Flares}
\begin{figure*}[htbp]
\begin{center}
 \includegraphics[scale=0.32]{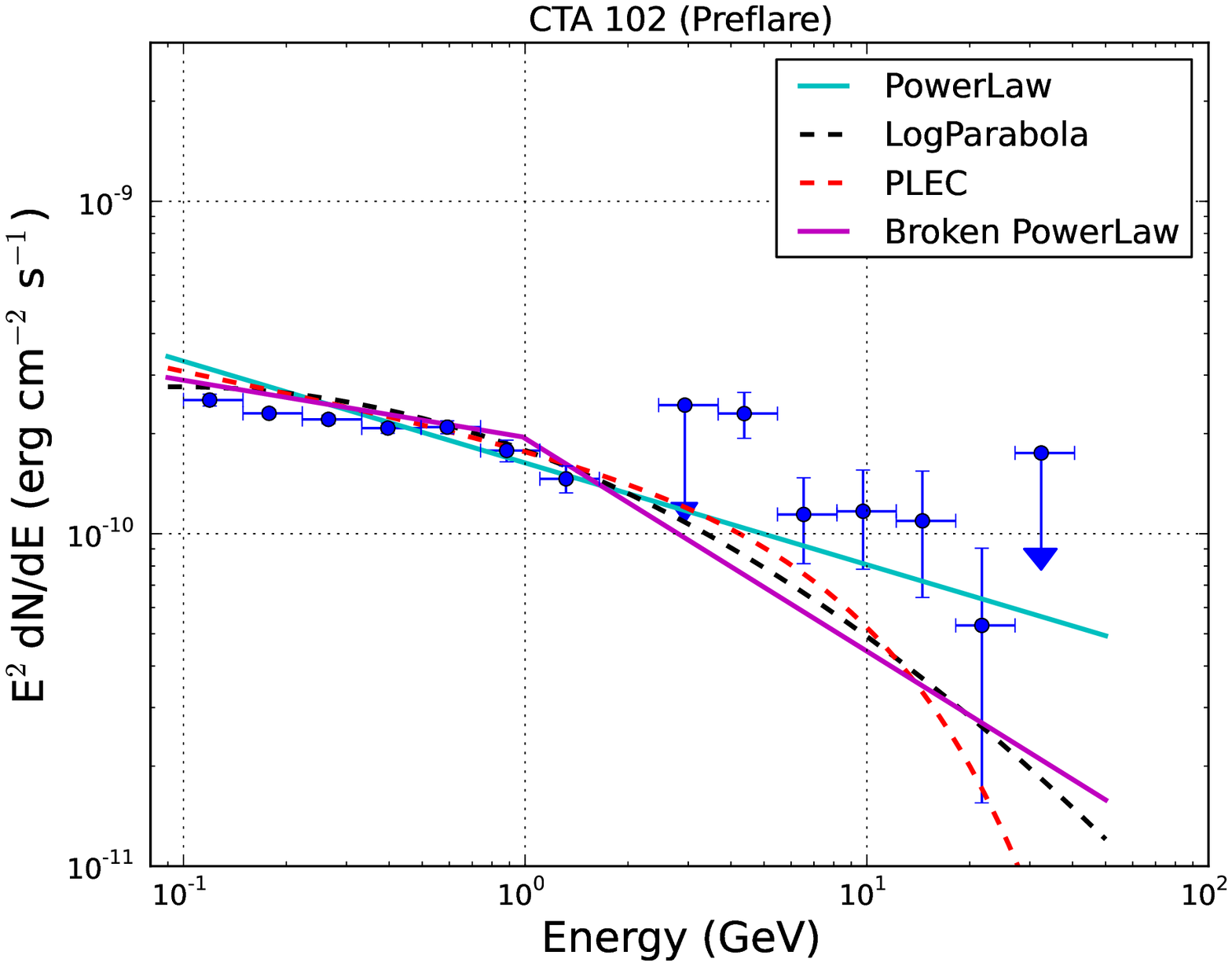}
 \includegraphics[scale=0.32]{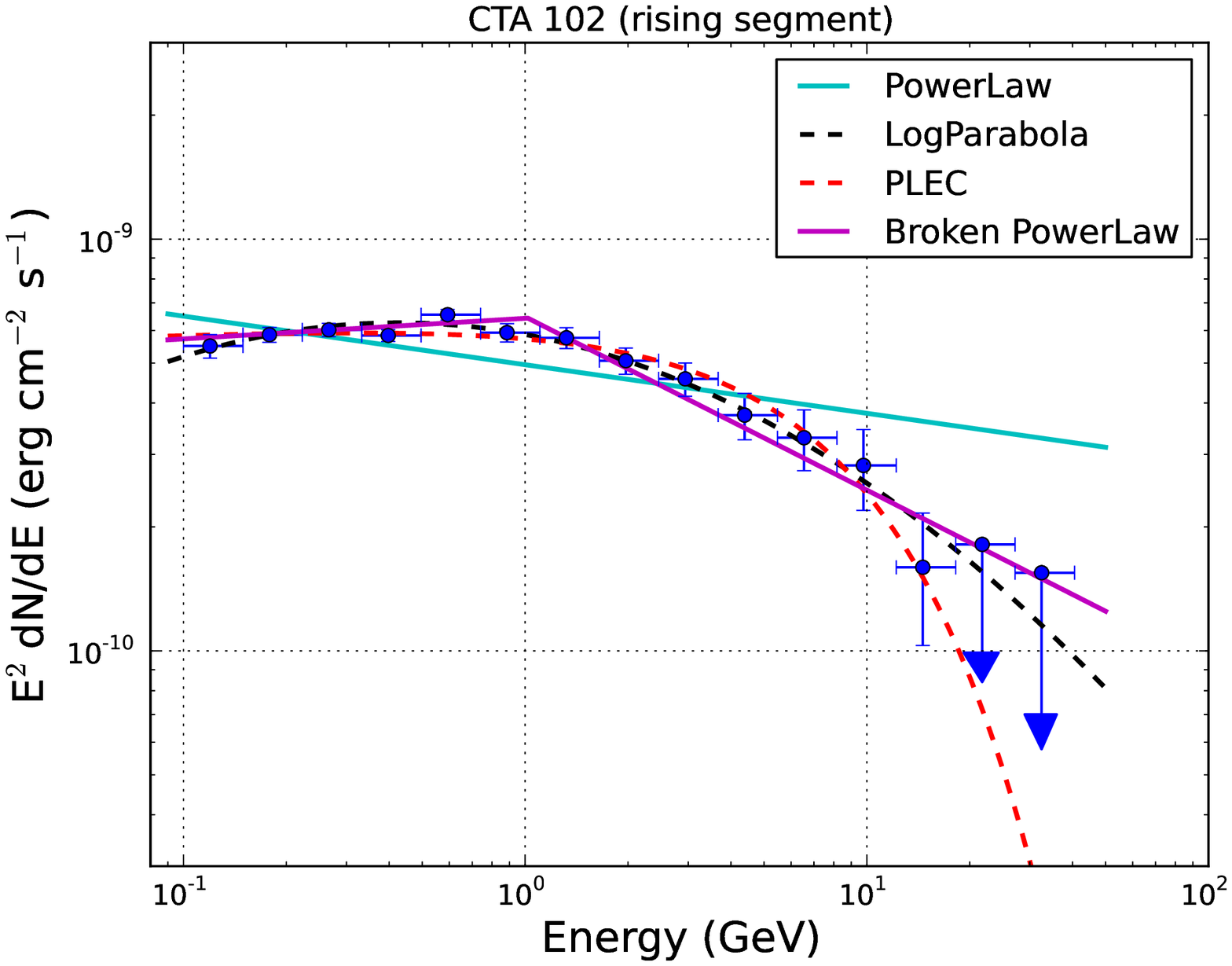}
\end{center}
\begin{center} 
 \includegraphics[scale=0.32]{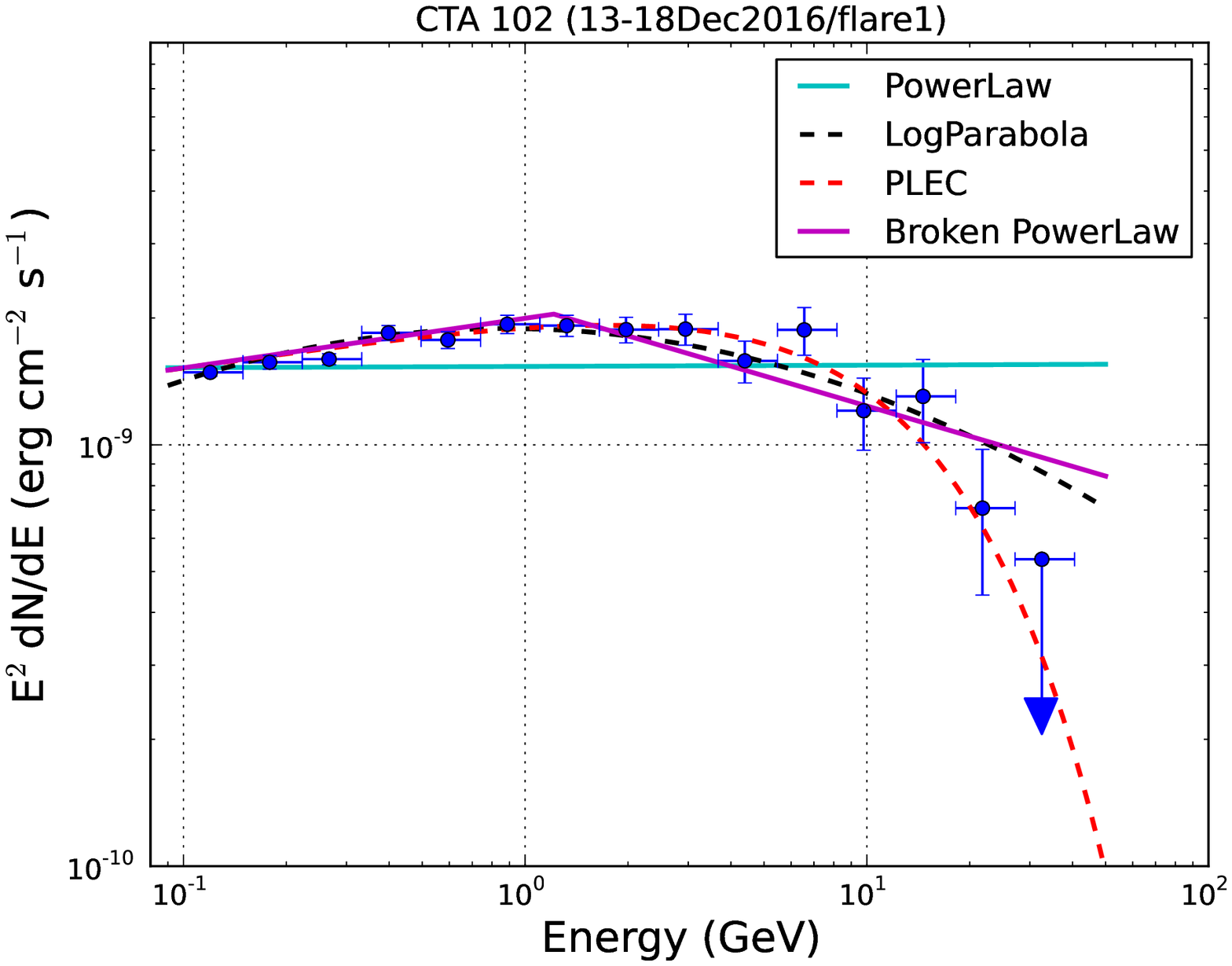}
 \includegraphics[scale=0.32]{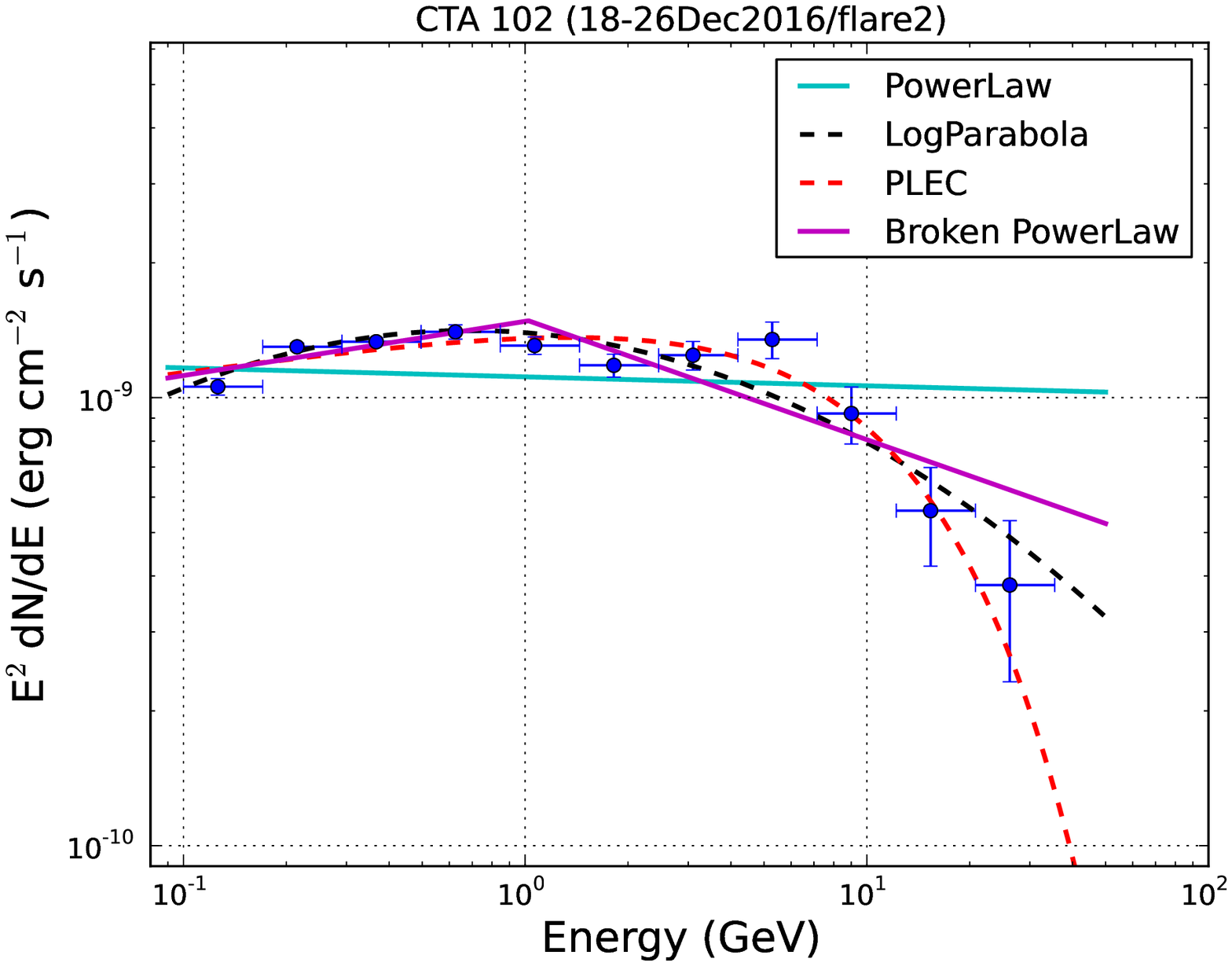}
 
\end{center}
\begin{center} 
 \includegraphics[scale=0.32]{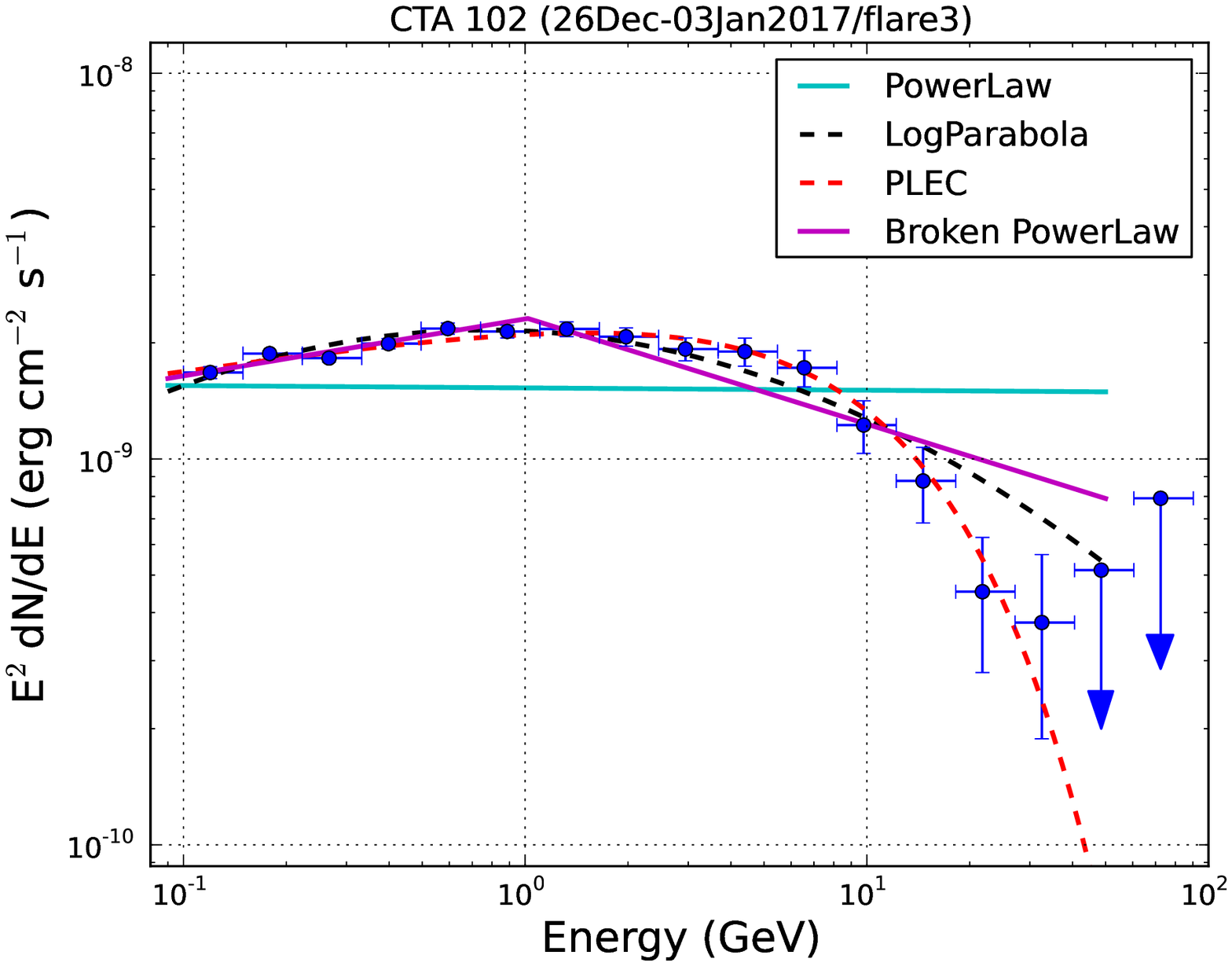}
 \includegraphics[scale=0.32]{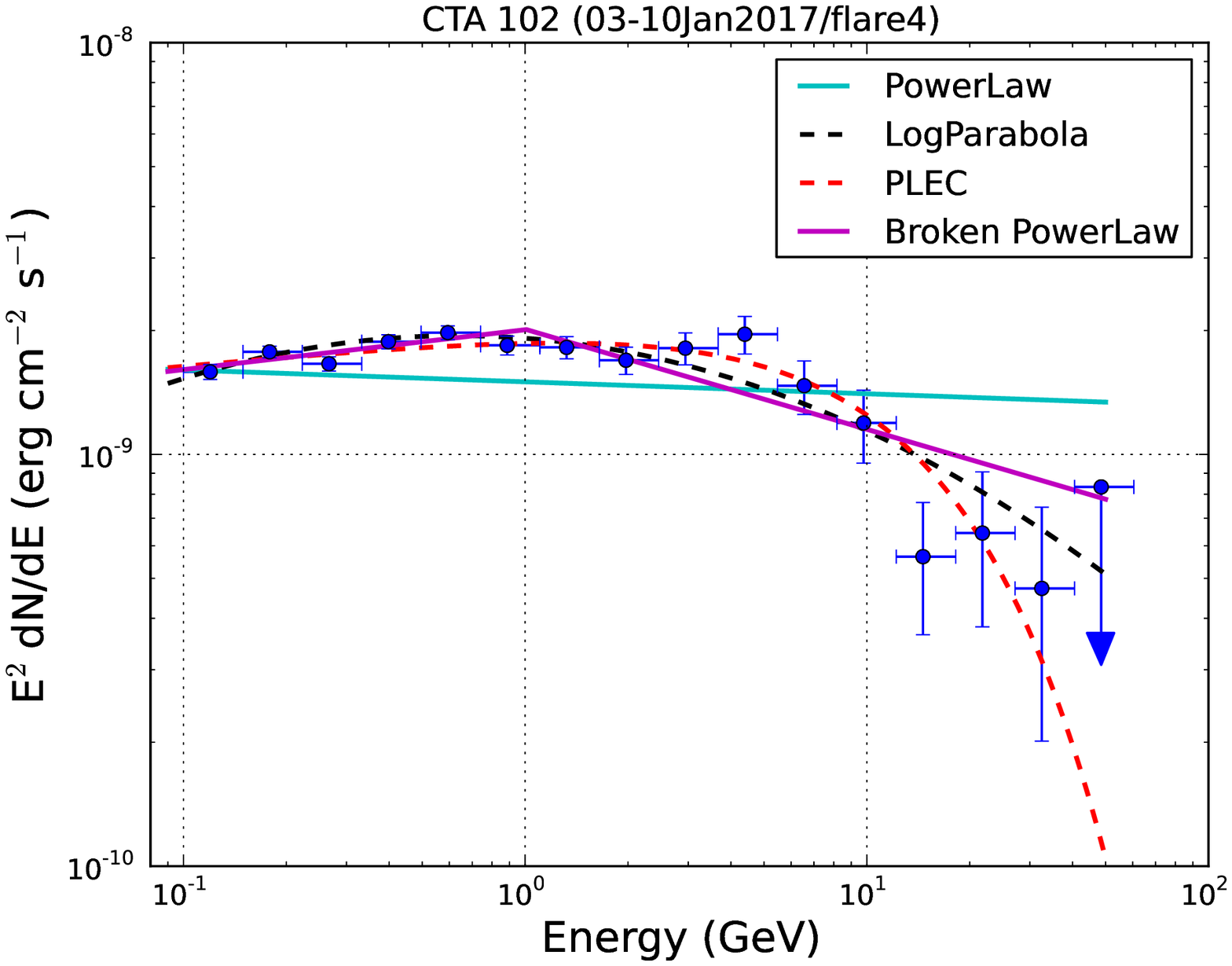}
 \includegraphics[scale=0.32]{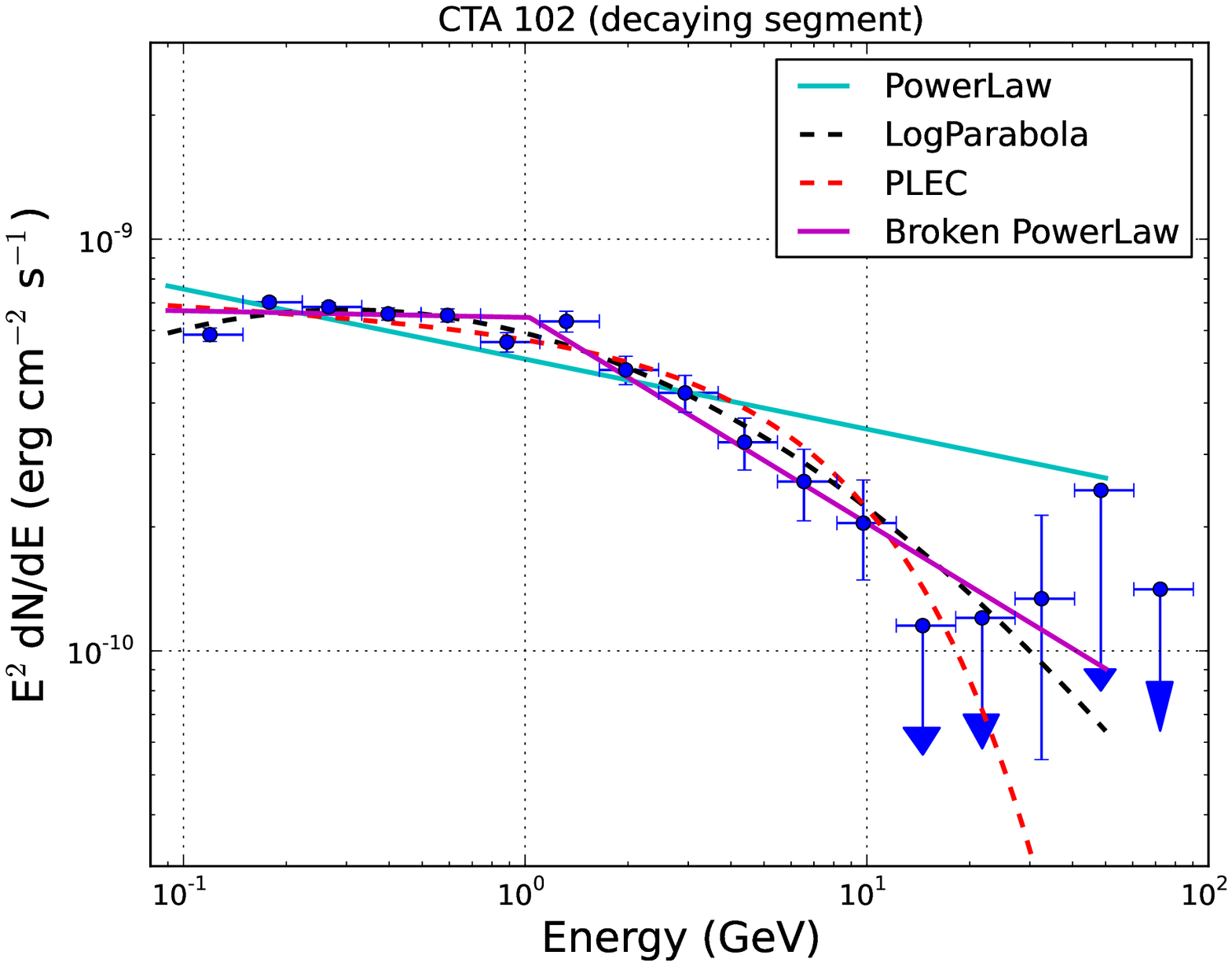}
\end{center}
\caption{Fermi-LAT SEDs during different activity periods shown in Figure \ref{fig:A}. The data points are fitted with four different
functional forms PL, LP, PLEC, and BPL shown in cyan, black, red, and magenta respectively. 
The fitted parameter values are displayed in Table \ref{Table:T3}.}
\label{fig:F}
\end{figure*}

In this section we have focused on the details of the $\gamma$-ray SEDs of Sep 2016-Mar 2017 flares. From the analysis we
have found the four flares accompanied by quiescent state (i.e. pre-flare), rising and decaying segments before and after the
flaring period. We have performed the spectral analysis of these phases separately.
Four different functions (PL, LP, PLEC and BPL) defined by Equations \ref{1}, \ref{2}, \ref{3}, and \ref{4} have been used to fit the spectral data points 
using the likelihood analysis. 
 The selection of these functional forms is motivated by earlier studies on spectral analysis of blazar flares (\citealt{Ackermann et al.(2010)}).
Along with the fitting parameters the likelihood analysis also returns the Log(likelihood) value which tells 
us about the quality of the 
unbinned fit. The spectral analysis for all the phases are shown in Figure \ref{fig:F} and their corresponding fitted parameters are 
presented in Table \ref{Table:T3}. The photon flux increases drastically during flares.  Moreover, significant spectral hardening 
is present in the photon spectra from flares. Flare-3 had the highest photon flux associated with maximum hardening in the spectrum.  
Similar features were also noted earlier from flares of other AGN like  3C 454.3 (\citealt{Britto et al. (2016)}).
The Spectral curvature has been identified with the TS$_{curve}$ which is defined as TS$_{curve}$ = 2(log $\mathcal{L}$(LP/PLEC/BPL) -- log $\mathcal{L}$(PL)) 
\citep{Nolan et al. (2012)}.
The spectral curvature is significant if TS$_{curve}>$ 16 \citep{Acero et al. (2015)}.
From the parameter values mentioned in Table \ref{Table:T3} maximum curvature in the $\gamma$-ray spectra has been noticed during rising, flaring and decaying states. 

 The \textit{Fermi}-LAT data from pre-flare is best fitted by a BPL or a LP function. The TS$_{curve}$ values do not differ much in these two cases. 
The rising segment is best fitted by a LP function. Other than flare-1 which gives a best fit to PLEC, all the three flares and the decaying 
segment are also best fitted by LP function. Flare-1 has comparable $TS_{curve}$ values for PLEC and BPL functions.
We infer from these results that the observed curvature in the gamma ray spectra during pre-flare, rising, flare and decaying states could be due to the
curvature in the spectrum of the relativistic electrons in the emission region.

\subsection{Modelling the SEDs}
We used the publicly available time dependent code GAMERA\footnote{http://joachimhahn.github.io/GAMERA} (\citealt{Hahn (2015)}) for modelling spectral 
energy distributions from astrophysical objects.
It solves the time dependent continuity equation to calculate the propagated electron spectrum. Subsequently the synchrotron and inverse Compton 
emissions for that electron spectrum are calculated. We have assumed a spherical emission region or blob moving relativistically to model the jet
emission. 

In Table \ref{Table:T4} our results of spectral fitting are displayed for the pre-flare, rising, flare and decaying states. 
In most cases log-parabola function gives the best fit to the \textit{Fermi}-LAT data.
A log-parabola photon spectrum can be produced by radiative losses of a log-parabola electron spectrum. 
The injected electron spectrum $Q(E,t)$ could be a log-parabola spectrum (\citealt{Massaro et al. (2004)}) if the probability of 
acceleration decreases with increasing energy. This electron spectrum becomes steeper after undergoing radiative losses which we denote below 
by $N(E,t)$.

The continuity equation for the electron spectrum in our study is
\begin{equation}\label{8}
\frac{\partial N(E,t)}{\partial t}=Q(E,t)-\frac{\partial}{\partial E}\Big(b(E,t) N(E,t)\Big)
\end{equation}
The energy loss rate of electrons due to synchrotron, 
Synchrotron Self-Compton and External Compton emission has been denoted by $b(E,t)$. GAMERA code calculates the inverse Compton emission using the full
Klein-Nishina  cross-section from \citet{Blumenthal and Gould (1970)}.
 
We have not included diffusive loss as it is assumed to be insignificant compared to the radiative losses by the electrons. 
The photons emitted from the broad line region (BLR) are the targets for external Compton emission. 
BLR photon density in the comoving/jet frame of Lorentz factor $\Gamma$ is   
\begin{equation} \label{5}
 U'{_{BLR}} = \frac{\Gamma^{2} {\eta_{BLR}} L_{disk}} {4 \pi c R_{BLR}^{2}}
\end{equation}
where the photon energy density in BLR is only a fraction $\eta_{BLR}\sim 0.02$(2\%) of the disc photon energy density. 
The BLR size is important to estimate the BLR energy density as well as absorption of $\gamma$-ray from the emission region. 
The accretion disk luminosity $L_{disk}$ = 3.8$\times$10$^{46}$
erg/s, central black hole mass $M_{BH}$ $\sim$ 8.5$\times$10$^{8}$ $M_{\odot}$ and Eddington luminosity $L_{Edd}$ = 1.1$\times$10$^{47}$ erg/s are 
estimated by \citet{Zamaninasab et al. (2014)}.  We have used their estimated value of $L_{disk}$ and $\Gamma=15$ to calculate $U'{_{BLR}}$. 
We have assumed the radius of the BLR region to be $R_{BLR}$ = 6.7$\times$10$^{17}$ cm following \citet{Pian et al. (2005)}. 

The accretion disk emission is also included in estimating the external Compton emission by the relativistic electrons in the jet.
The energy density in the comoving frame (\citealt{Dermer and Menon (2009)}) is calculated from the following equation
\begin{equation}\label{9}
U'_{disk}=\frac{0.207 R_g l_{Edd} L_{Edd}}{\pi c z^3 \Gamma^2}
\end{equation}
The gravitaional radius is denoted by $R_g$, the Eddington ratio by $l_{Edd}$ = L$_{disk}$/L$_{Edd}$  and $z=6.7\times 10^{17}$cm is the distance 
of the emission region from the black hole. The accretion disk temperature is estimated from \citet{Dermer and Menon (2009)} by using the 
$l_{Edd}$ and the mass of the central black hole ($M_{BH}$).
We note that the external Compton emission by  NIR/optical/UV photons emitted by disk  and dusty torus based clouds irradiated by a spine-sheath
jet could be important  (\citealt{Finke (2016)}, \citealt{Gaur et al. (2018)}, \citealt{Breiding et al. (2018)}) in some cases.
Due to a lack of observational evidence we are not including dusty torus as target photon field in our model.

The magnetic field inside the blob, Doppler factor of the blob, spectral indices of the injected electron spectrum, luminosity in injected 
electrons are the model parameters, whose values are optimized to fit the SEDs in Figure \ref{fig:G}.  
\subsection{Multi-wavelength SEDs}
Time dependent multi-wavelength modelling has been done with Swift UV, X-ray and \textit{Fermi}-LAT $\gamma$-ray data for the pre-flare, 
rising segment, four flares and decaying segment.

\begin{figure*}%
\centering 
\subfigure[]{%
\label{fig:ex3-i}%
\includegraphics[scale=0.40]{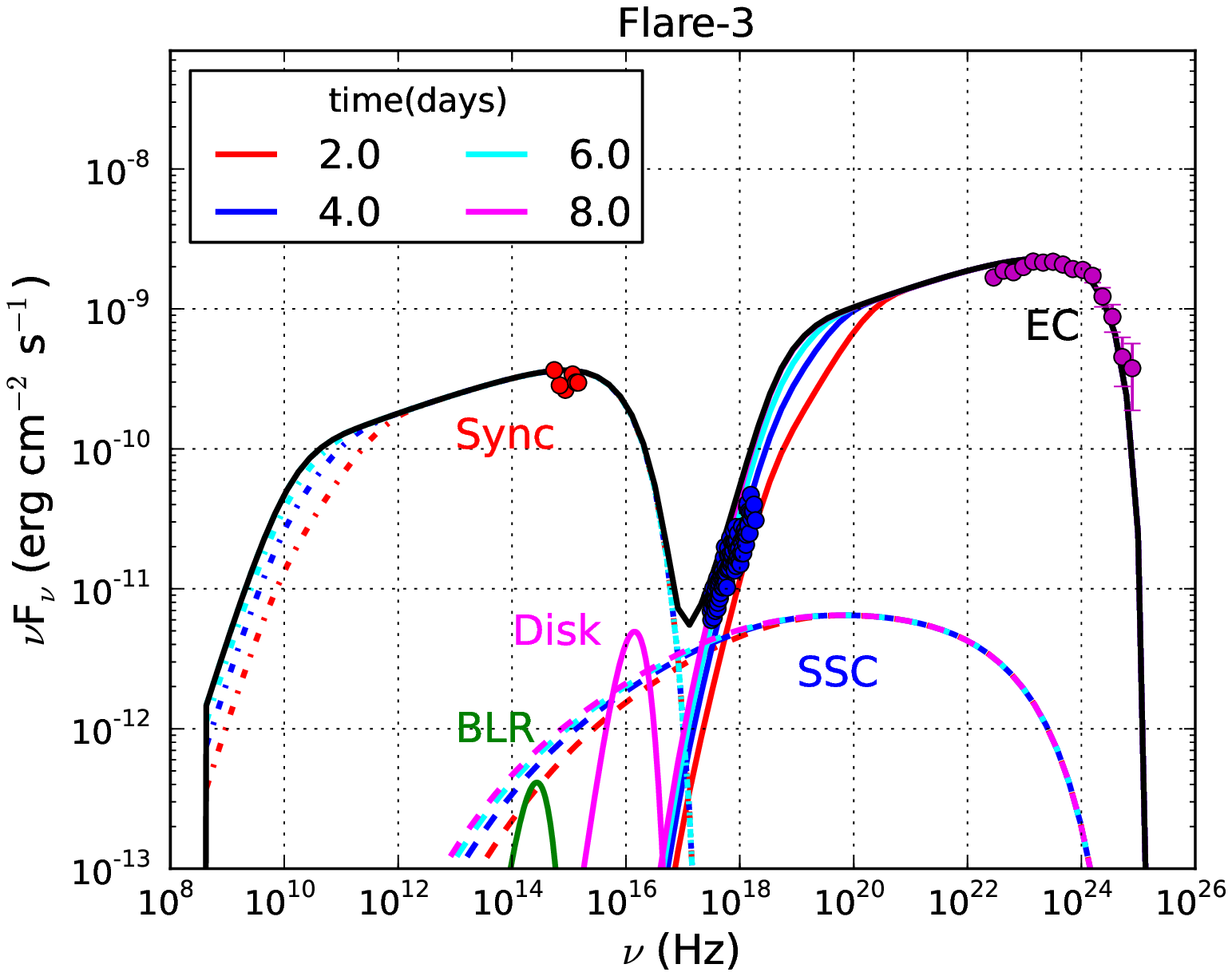}}%
\subfigure[]{%
\label{fig:ex3-j}%
\includegraphics[scale=0.40]{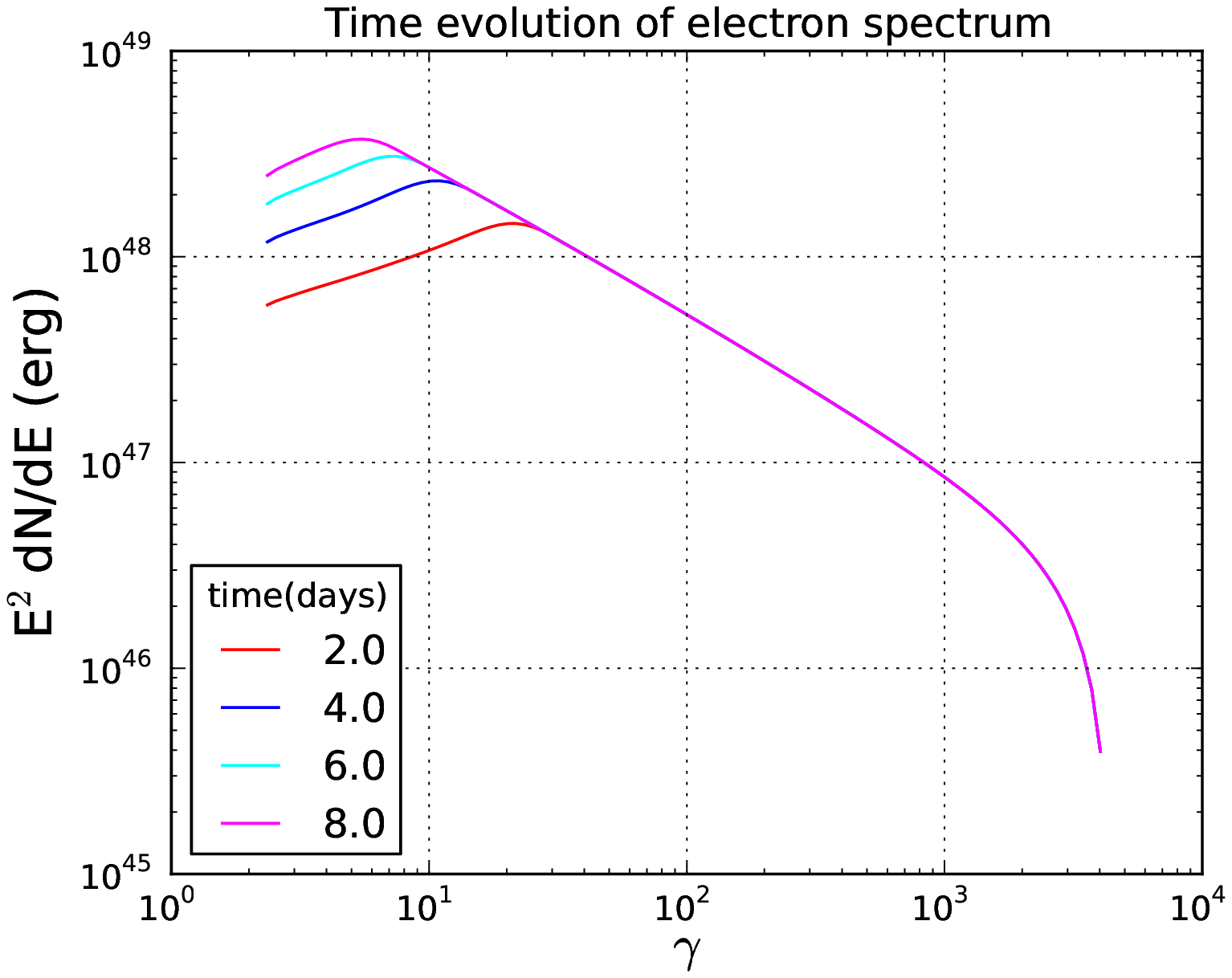}}%
\qquad  
\centering 
\subfigure[]{%
\label{fig:ex3-k}%
\includegraphics[scale=0.40]{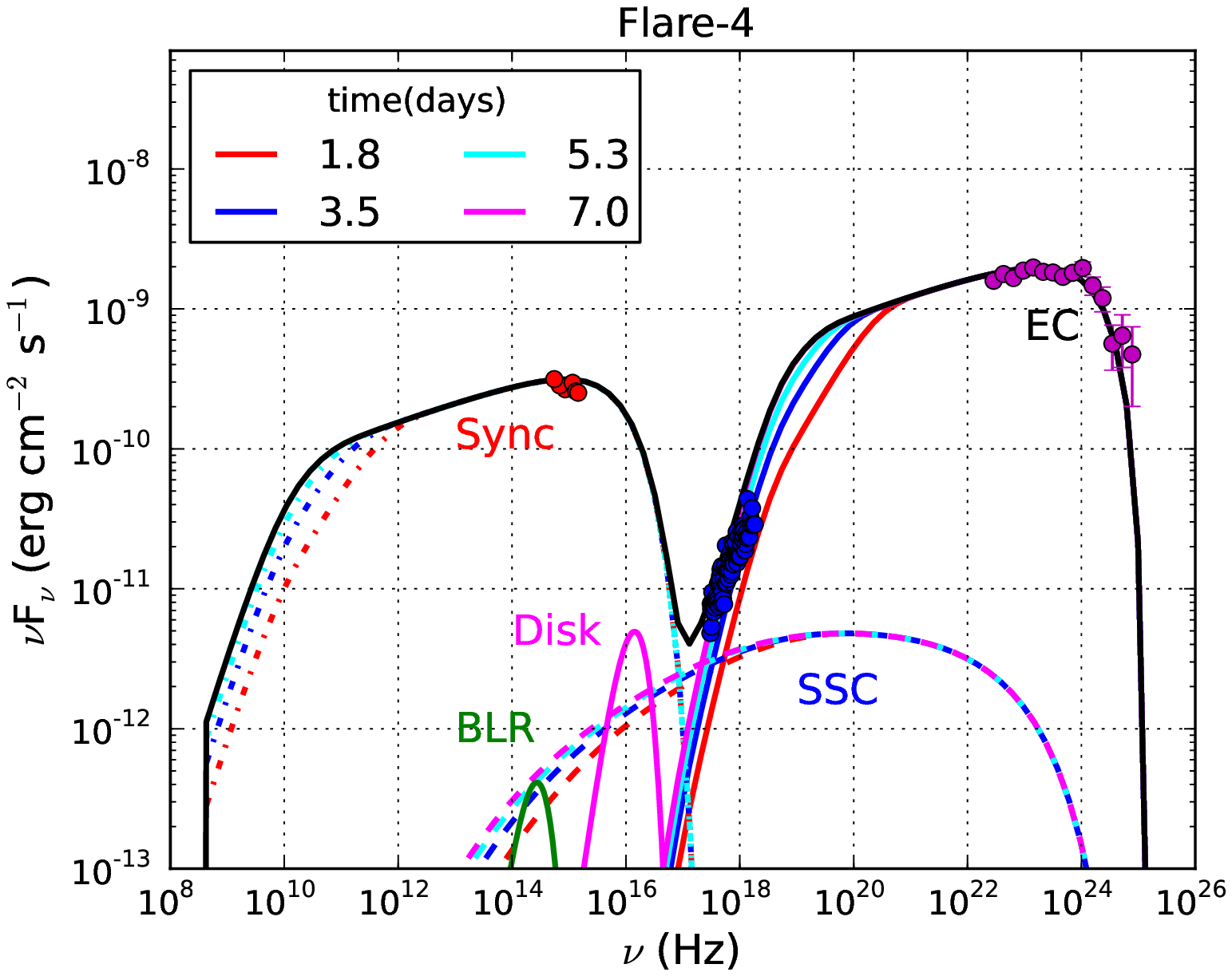}}%
\subfigure[]{%
\label{fig:ex3-l}%
\includegraphics[scale=0.40]{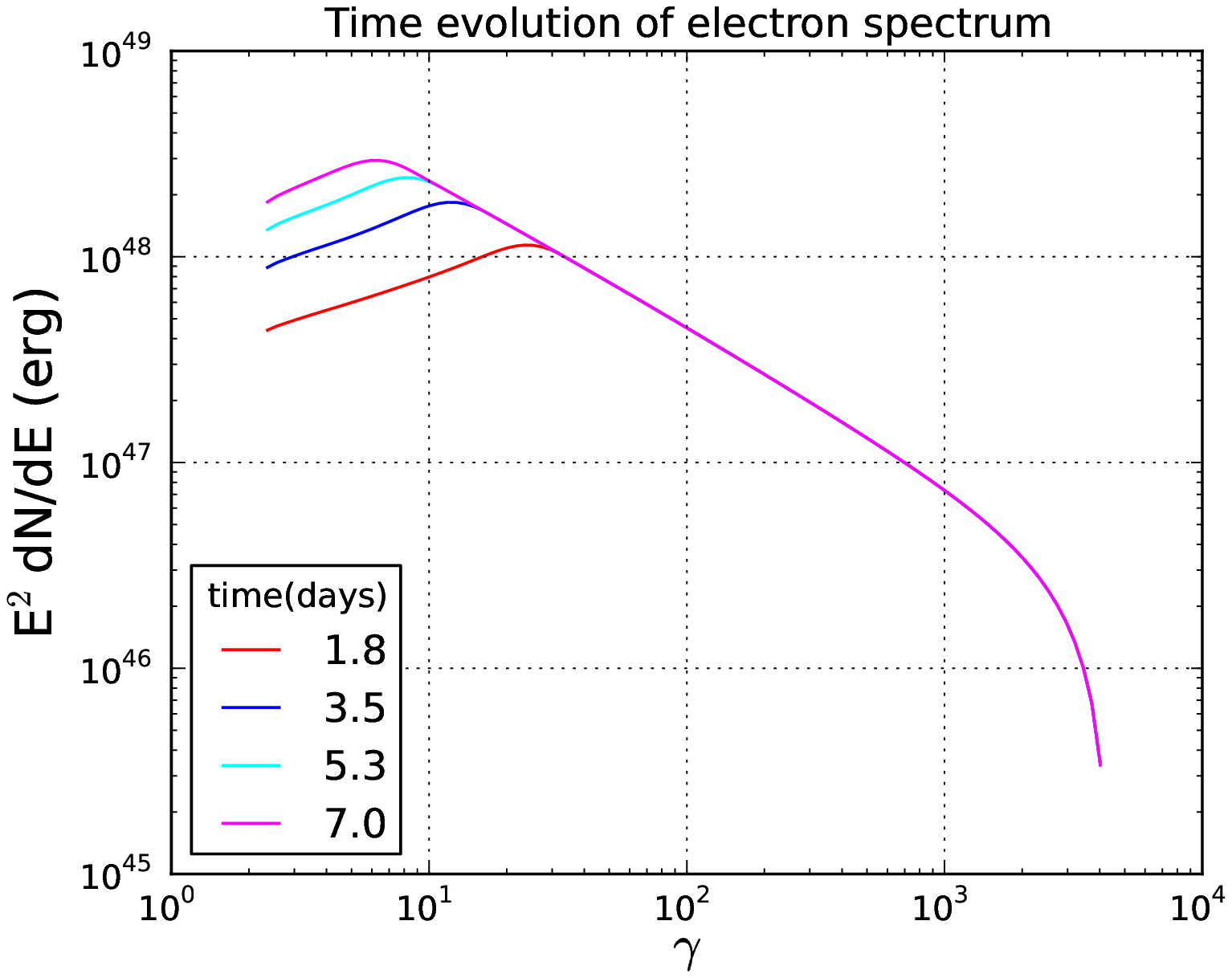}}%
\qquad   
\centering 
\subfigure[]{%
\label{fig:ex3-m}%
\includegraphics[scale=0.40]{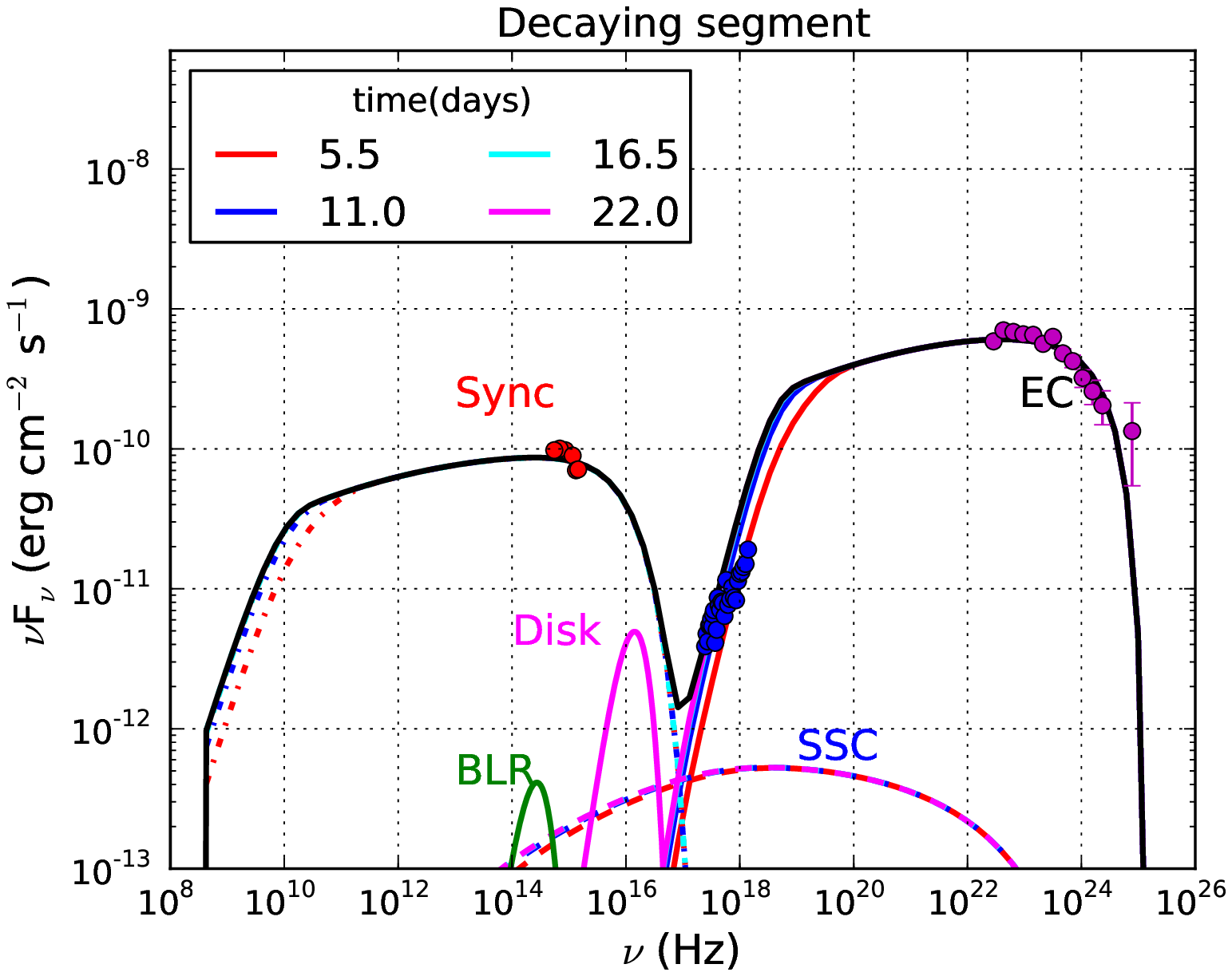}}%
\subfigure[]{%
\label{fig:ex3-n}%
\includegraphics[scale=0.40]{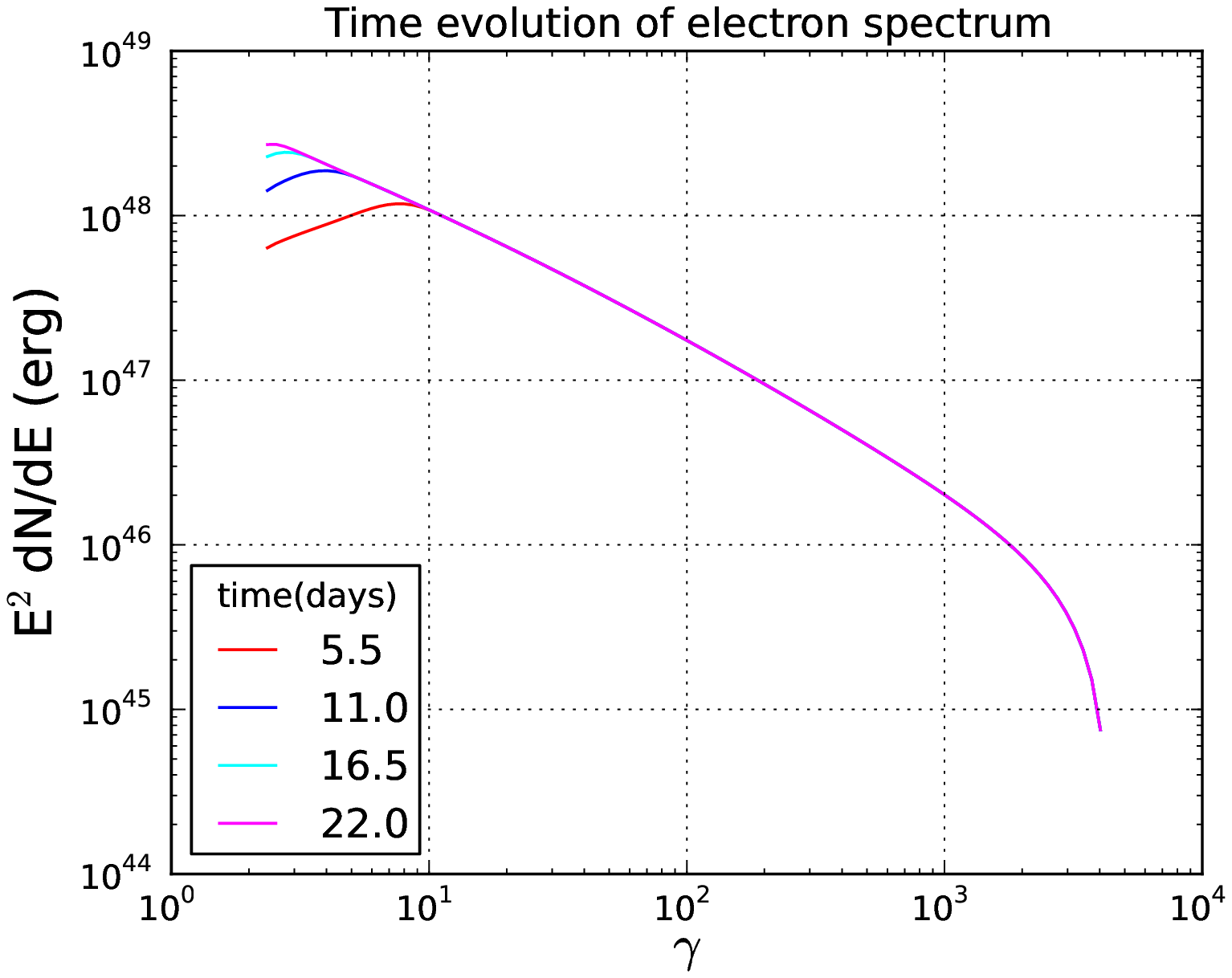}}%
\caption{Modelled multi-wavelength SEDs during different activity periods shown in the left panel. The plots are arranged in the following order pre-flare, rising segment, flare-1, flare-2, flare-3, flare-4 and decaying segment. Time evolution of electron spectra is  shown in the right panel. Each activity period is divided in four equal time intervals and shown in different colours. The model parameters are mentioned in Table \ref{Table:T4}. }

\label{fig:G}
\end{figure*}

In each phase the injected electron spectrum 
evolves with time as the electrons lose energy radiatively. The time evolution of the electron spectra is shown in right panel of Figure \ref{fig:G}.
 The Doppler factor is assumed to be $35$, which is high compared to the values estimated by \citet{Casadio et al. (2015)} and Lorentz 
factor $15$. The size of the emission region is adjusted to $6.5\times 10^{16}$cm so that SSC emission is not too high.
In Figure \ref{fig:G} the X-ray data constrains the SSC emission.
We note that intra-night variability observed in optical flux suggests an upper limit on the size of the emission region 
$4.5\times10^{16}$cm for Doppler factor $35$, which is comparable to the size used in our model. In \textit{Fermi}-LAT data the variability time is observed
to vary in the range of one hour to several days. The magnetic field rises from 4.0 Gauss to 4.2 Gauss and the luminosity injected in electrons
increases nearly by a factor of seventy as the source transits from the pre-flare to the flaring state. We also note that most of the jet power is
in the magnetic field not the injected electrons. The maximum jet power required in our model is $6.6\times 10^{46}$ erg/sec.

We have calculated the photon flux during each phase (pre-flare, rising segment, flare-1, flare-2, flare-3, flare-4 and decaying segment) 
and compared with the data in Figure \ref{fig:G}. The values of the parameters, displayed in Table \ref{Table:T4}, 
are the best model parameters to fit the observed photon flux.
 \section{Discussion}
Sep 2016 to Mar 2017 was the active period for the blazar CTA 102, not only in $\gamma$-ray but also in X-ray and optical/UV. In these
190 days CTA 102 had four major $\gamma$-ray flares with the highest flux of 30.12$\pm$4.48 (for 90 minutes binning).
CTA 102 was also very bright in X-ray and optical/UV. All the $\gamma$-ray flares were  simultaneous with the flares in X-ray 
and optical/UV, as shown in Figure \ref{fig:A}. 
  
The time evolution of the SEDs are shown in Figure \ref{fig:G}.

 The analysis of \textit{Fermi}-LAT data shows variability in gamma ray data in time scale of an hour to several days.
Intra-night variability has been observed in the optical flux (\citealt{Bachev et al. (2017)}). For Doppler factor 35 intra-night variability 
time scale gives an upper limit of $4.5\times 10^{16}$ cm on the size of the emission region. We have used $6.5\times 10^{16}$ cm in our work 
so that the SSC emission is not too high.
The region size is related to the Doppler factor and variability time scale as
 \begin{equation}
 R \leq c\, t_{var} \, \delta \, (1+z)^{-1}
 \end{equation}
It is important to note that the above relation is an approximation and there are other effects which may introduce large errors in determining
the size of the emission region \citep{Protheroe (2002)}.

We note that there could also be EC emission from the target photons in the dusty torus
region, however due to a lack of observational information we do not include the dusty torus region in our model.

The values of the parameters fitted in our multi-wavelength
modelling are shown in Table \ref{Table:T4}. 

 The rise in injected luminosity of electrons or jet power causing the rise in multi-wavelength emission from the jet of CTA 102 during  the flaring state 
can be explained with increase in accretion rate of the super massive black hole which powers the jet.  
The relationship between jet power and accretion in blazars has been well studied earlier. 
A large sample of blazars was used to study the jet-disc connection by \citet{Sbarrato et al. (2014)}. They noted that BLR luminosity is a tracer of accretion rate while
gamma ray luminosity is tracer of jet power. It was found that the two luminosities are linearly connected.

Fluctuation in luminosity and variability in AGN was discussed as a stochastic process in \citet{Kelly et al. (2011)}.
They gave a relation between characteristic time scale of high frequency X-ray emission and  black hole mass of AGN.

Variability in blazar emission on time scale of days to years due to change in accretion rate was  also discussed by \citet{Sartori et al. (2018)}.   
They  modelled AGN variability  as a result of variations in fuelling of super massive black hole following the idea of \citet{Kelly et al. (2011)}.
Unsteady fuelling of black hole may occur due to physical processes of different spatial scales. 
Disc properties like its structure, viscosity and the system's response to perturbations could be one of the factors influencing the conversion of 
gravitational energy to jet luminosity
 \citep{Shakura and Sunyeav (1973)}.
 
Possible accretion disk origin of variability in jet of Mrk 421, which is a BL Lac, has been reported by \citet{Chatterjee et al. (2018)}.
This source  having a weak disc emission  and strong jet emission in X-rays showed  a break in  power spectral density  which could be connected to variation
in accretion rate.
This strengthens the motivation for accretion-jet  scenario of blazar flares.

Here we discuss about the other models considered earlier to explain flares of CTA 102. The evolution of physical parameters during the historical radio outburst
in April 2006  was studied with shock-in-jet model  \citep{Marscher and Gear(1985)} by  \citet{Fromm et al. (2011)}. In this model a travelling 
shock wave evolves in a steady state jet.
During the passage of the shock through the jet  the relativistic electrons carried down by the shock front  get energy while crossing the shock front.
Accelerated electrons cool down within a small layer behind the shock front. 
Magnetic field, Doppler factor, spectral index of electron spectrum and its normalisation constant, also the region size evolve as a power law in distance along the jet.
Dominant loss  energy mechanism of  electrons was Compton during the first stage of the flare and adiabatic during the final stage. They concluded that a change in 
the evolution of the Doppler factor can not explain the observed temporal evolution of  the turnover  frequency and turnover flux  density.
They suggested shock-shock interaction \citep{Fromm et al. (2013a)}  between travelling and standing shock wave might be the possible mechanism 
as it provides a better understanding of the evolution  of the physical parameters compared to the shock-in-jet model.

\citet{Fromm et al. (2013a)} found variability Doppler factor decreases from 17  in region C to 8 in region D with distance along the jet. At the 
same time apparent speed decreases and the viewing angle increases. The deceleration of plasma flow along the jet was inferred  from their 
analysis. The stationary behaviour near $r\sim 1.5$ mas could be from recollimation shock at that position. The increase in viewing angle from 
region C to region D could be from helical instabilities due to asymmetric pressure in the jet.

In September - October 2012 an exceptional outburst of CTA 102 was recorded \citep{Larionov et al. (2016)}. The multi-wavelength data covering 
near infrared to gamma ray frequencies was modelled assuming a radiating blob or shock wave moves along a helical path down the jet. The changes
in the viewing angle  caused by motion of shock wave along the helical path down the jet implied large changes in the value of the Doppler factor
from 28 to 16. They inferred co-spatiality of optical and gamma ray emission regions which supports SSC mechanism of emission.

The gamma ray flare of January 2016 was studied by a helical jet model by \citet{Li et al. (2018)}. They inferred a Doppler factor of 17.5 and
size of the emission region 0.11-0.32 pc.
They further inferred that the emission region is located  at a distance of 5.7 to 16.7 pc from the central engine assuming a conical jet 
geometry. At a distance of 1 pc along the jet they estimated a magnetic field  1.57 G using the core shift method.

In the paper by \citet{Zacharias et al. (2017)} the flare of CTA 102 in 2016 and 2017 has been modelled by ablation of  a gas cloud by a 
relativistic jet. They have assumed  that gradual increase in number of injected electrons in the jet during the flare is due to slice by slice
ablation of the cloud, until the centre of the cloud is reached. 
Subsequently the particle injection decreases which results in decay of the flare. The value of Doppler factor and BLR temperature used in our model is similar to
\citet{Zacharias et al. (2017)}. They have assumed a magnetic field 3.7 G which is comparable to the value assumed in our work. The region size 
is smaller $2.5\times 10^{16}$ cm in their model. They have assumed time dependent luminosity and spectral index of injected electrons. 
In our case these variables are constants and adjusted for each state to obtain good fit to the data.
In their model EC by BLR photons is the main radiative loss mechanism of relativistic electrons and SSC emission is always insignificant.

The multi-wavelength emission from CTA 102 has also been analysed by \citet{Gasparyan et al. (2018)}. They have considered very short periods or
time intervals of observation and the data is fitted with SSC and EC by BLR and torus photons. They noted spectral curvature and hardening in 
the gamma ray spectra. In their work the magnetic field and the luminosity in injected electrons differ significantly from one epoch to another
epoch.
Thus the high activity states of CTA 102 have been analysed and modelled earlier in different ways to obtain good fit to the observed data and 
the variations in the values of the physical parameters (magnetic field, luminosity in injected electrons) are model dependent.

A study of the time evolution of the physical parameters (e.g. magnetic field, Doppler factor, spectral index and luminosity in electrons, 
region size) required for SED modelling in pre-flare, flare and decaying states is necessary 
 as the variations in the values of these parameters could be good indicators 
of the  underlying model. At least some of the models could be excluded in this way.
Simulated SEDs could be compared with the parametric fitting of SEDs for this purpose.

\section{Concluding Remarks}
We have studied the brightest flaring state of CTA 102 observed during Sep 2016 to Mar 2017. 
Four major flares have been identified 
between MJD 57735 -- 57763 in $\gamma$-ray. Similar flares have also been observed in X-rays and 
optical/UV frequencies during this period.   
The pre-flare, rising phase before 
the four consecutive flares, the four major flares and the decay phase at the end have all been analysed by spectral analysis of gamma ray data 
and multi-wavelength SED modelling. 
The highest energy photon detected during the flaring episode is $\sim$ 73 GeV. 
The multi-wavelength data has been modelled using the time 
dependent code GAMERA to estimate the synchrotron and inverse Compton emission. Our study shows that the data can be fitted by
a single zone model during various phases by varying the luminosity in injected electrons, slightly changing their spectral index  and the magnetic field.

\section{Acknowledgement}
The authors thank the referee for helpful comments to improve the paper.
This work has made use of public \textit{Fermi} data obtained from the Fermi Science Support Center (FSSC), provided by NASA Goddard Space Flight
center. This research has also made use of data, software/tools obtained from NASA High Energy Astrophysics Science Archive Research Center 
(HEASARC) developed by Smithsonian astrophysical Observatory (SAO)
and the XRT Data Analysis Software (XRTDAS) developed by ASI Science Data Center, Italy. 

\begin{table*}[htbp]
\caption{Results of temporal fitting with sum of exponentials (Equation \ref{6} in the text) for different peaks of the flares.
Column 2 represents the time (in MJD) at which the peaks were observed and the peak fluxes are given in column 3. 
The fitted rise ($T_r$) and decay ($T_d$) times are mentioned in columns 4 \& 5}
\centering
\begin{tabular}{c c c c c}

\hline
\\
&& flare-1&& \\
\\
Peak  & $t_0$  &$F_0$&  $T_r$ & $T_d$  \\[0.5ex]
\\
     &(MJD) & ($10^{-6}$ ph cm$^{-2}$ s$^{-1}$) & (hr) & (hr)  \\
\\
\hline
\\
P1  & 57736.4 & 12.81$\pm$1.42 & 7.66$\pm$1.35 & 7.76$\pm$1.36  \\
\\
P2  & 57738.4 & 19.45$\pm$0.68 & 10.01$\pm$0.92 & 6.51$\pm$0.59  \\
\\
\hline
\\
&& flare-2&& \\
\\
\hline
\\
P1  & 57741.6 & 9.12$\pm$0.55 & 13.21$\pm$2.45 & 6.31$\pm$1.70  \\
\\
P2  & 57742.6 & 8.70$\pm$0.56 & 4.01$\pm$1.15 & 9.48$\pm$2.11  \\
\\
P3  & 57744.6 & 14.38$\pm$1.24 & 12.50$\pm$1.84 & 15.30$\pm$3.35  \\
\\
P4  & 57746.6 & 10.84$\pm$1.46 & 14.53$\pm$4.37 & 9.83$\pm$2.09  \\
\\
\hline
\\
&& flare-3&& \\
\\
\hline
\\
P1  & 57750.1 & 20.93$\pm$1.24 & 7.20$\pm$1.00 & 7.45$\pm$1.91  \\
\\
P2  & 57750.9 & 21.17$\pm$1.67 & 5.61$\pm$1.62 & 6.06$\pm$1.84  \\
\\
P3  & 57751.6 & 20.09$\pm$1.60 & 4.64$\pm$1.61 & 4.56$\pm$0.95  \\
\\
P4  & 57752.6 & 20.82$\pm$1.08 & 5.05$\pm$0.85 & 11.41$\pm$1.50  \\
\\
P5  & 57753.6 & 14.79$\pm$1.01 & 4.94$\pm$1.41 & 4.49$\pm$1.17  \\
\\
\hline
\\
&& flare-4&& \\
\\
\hline
\\
P1  & 57758.1 & 13.01$\pm$1.20 & 7.21$\pm$1.66 & 4.89$\pm$1.41  \\
\\
P2  & 57759.6 & 22.50$\pm$1.73 & 10.07$\pm$1.36 & 1.74$\pm$0.99  \\
\\
P3  & 57760.4 & 21.36$\pm$1.52 & 8.97$\pm$1.21 & 8.72$\pm$0.82  \\
\\
\hline
\end{tabular}
\label{Table:T2}
\end{table*}

\begin{table*}[htbp] 
\caption{ Results of SEDs fitted with different spectral distributions PL, LP, PLEC, and BPL.
The different states during the observations are mentioned in the 1st column. The values of the fitted fluxes and spectral indices are presented in  
columns 2 \& 3. TS$_{curve}$ = 2(log $\mathcal{L}$(LP/PLEC/BPL) -- log $\mathcal{L}$(PL)).}

\centering
\begin{tabular}{c c c c c c c}
\hline
\\
&&PowerLaw (PL)&&&&  \\
\\
Activity  &  F$_{0.1-300}$ $_{\rm{GeV}} $ & $\Gamma$ & && TS &  \\[0.5ex]
\\
& ($10^{-6}$ ph cm$^{-2}$ s$^{-1}$)&&&&  \\
\\
\hline
\\
pre-flare    & 1.44$\pm$0.04 & 2.31$\pm$0.02 & - && 6965.29 & - \\

rising segment   & 3.87$\pm$0.07 & 2.12$\pm$0.01 & - && 18236.04 & - \\

flare-1    & 10.60$\pm$0.17 & 1.99$\pm$0.01 & - && 26338.38 & -  \\

flare-2    & 7.87$\pm$0.12 & 2.02$\pm$0.01 & - && 30981.15 & - \\

flare-3   & 11.90$\pm$0.14 & 2.01$\pm$0.01 & - && 46205.62 & - \\

flare-4    & 11.10$\pm$0.18 & 2.03$\pm$0.01 & - && 24250.71 & - \\

decaying segment   &4.17$\pm$0.07 & 2.17$\pm$0.01 & - && 18320.72 & -  \\
\\
\hline
\\
&&LogParabola (LP)&&&  \\
\\
Activity  &  F$_{0.1-300}$ $_{\rm{GeV}} $ & $\alpha$ & $\beta$& &  TS & TS$_{curve}$ \\[0.5ex]
\\
& ($10^{-6}$ ph cm$^{-2}$ s$^{-1}$) &&&&  \\
\\
\hline
\\
pre-flare    & 1.38$\pm$0.04 & 2.18$\pm$0.03 & 0.08$\pm$0.02  && 6980.74 & 15.45 \\

rising segment  & 3.66$\pm$0.07 & 1.96$\pm$0.02 & 0.09$\pm$0.01  && 18333.51 & 97.47 \\

flare-1    & 10.30$\pm$0.17 & 1.87$\pm$0.02 & 0.06$\pm$0.01 && 26408.49 & 70.11  \\

flare-2    & 7.60$\pm$0.12 & 1.87$\pm$0.02 & 0.08$\pm$0.01 && 31083.27 & 102.12 \\

flare-3   & 11.50$\pm$0.14 & 1.85$\pm$0.02 & 0.08$\pm$0.01 && 46437.84 & 232.22 \\

flare-4    &10.70$\pm$0.19 & 1.89$\pm$0.02 & 0.07$\pm$0.01 && 24332.05 & 81.34 \\

decaying segment  &3.96$\pm$0.07 & 2.01$\pm$0.02 & 0.09$\pm$0.01 && 18412.52 & 91.80  \\
\\
\hline
\\
&&PLExpCutoff (PLEC)&&&  \\
\\
Activity  &  F$_{0.1-300}$ $_{\rm{GeV}} $ &$\Gamma_{PLEC}$ & E$_{cutoff}$ && TS & TS$_{curve}$  \\
\\
&($10^{-6}$ ph cm$^{-2}$ s$^{-1}$)&& [GeV] &&  \\
\\
\hline
\\
pre-flare    & 1.40$\pm$0.04 & 2.21$\pm$0.04 & 12.39$\pm$4.16 && 6976.88 &11.59 \\

rising segment  & 3.72$\pm$0.07 & 1.97$\pm$0.02 & 9.45$\pm$1.46 && 18326.98 & 90.94 \\

flare-1    & 10.40$\pm$0.17 & 1.88$\pm$0.02 & 14.21$\pm$2.22 && 26421.66 & 83.28  \\

flare-2    & 7.67$\pm$0.12 & 1.89$\pm$0.02 & 12.71$\pm$1.79 && 31081.92 & 100.77 \\

flare-3   & 11.60$\pm$0.14 & 1.87$\pm$0.02 & 11.98$\pm$1.33 && 46414.40 & 208.78 \\

flare-4    & 10.80$\pm$0.18 & 1.92$\pm$0.02 & 15.46$\pm$2.69 && 24326.29 & 75.58 \\

decaying segment   & 4.05$\pm$0.07 & 2.05$\pm$0.02 & 10.76$\pm$2.13 && 18385.50 & 64.78  \\
\\
\hline
\\
&&Broken PowerLaw (BPL)&&&  \\
\\
Activity  &  F$_{0.1-300}$ $_{\rm{GeV}} $ &$\Gamma_{1}$ & $\Gamma_{2}$ & E$_{break}$ & TS & TS$_{curve}$  \\
\\
&($10^{-6}$ ph cm$^{-2}$ s$^{-1}$)&&& [GeV] &&  \\
\\
\hline
\\
pre-flare    & 1.38$\pm$0.06 & 2.17$\pm$0.07 & 2.64$\pm$0.10 & 0.98$\pm$0.12 & 6984.17 & 18.88 \\

rising segment  & 3.69$\pm$0.09 & 1.96$\pm$0.03 & 2.42$\pm$0.05 & 1.02$\pm$0.09 & 18323.99 & 87.95\\

flare-1    & 10.40$\pm$0.17 & 1.88$\pm$0.02 & 2.24$\pm$0.04 & 1.21$\pm$0.11 & 26396.25 & 57.87  \\

flare-2    & 7.65$\pm$0.12 & 1.88$\pm$0.02 & 2.27$\pm$0.03 & 1.02$\pm$0.04 & 31060.34 & 79.19 \\

flare-3   & 11.60$\pm$0.14 & 1.85$\pm$0.02 & 2.28$\pm$0.03 & 1.01$\pm$0.14 & 46402.10 & 196.48 \\

flare-4    & 10.80$\pm$0.18 & 1.90$\pm$0.03 & 2.24$\pm$0.04 & 1.01$\pm$0.19 & 24312.09 & 61.38 \\

decaying segment   &4.00$\pm$0.07 & 2.02$\pm$0.03 & 2.51$\pm$0.06 & 1.03$\pm$0.14 & 18403.22 & 82.50  \\
\\
\hline
\end{tabular}
\label{Table:T3}
\end{table*}

\begin{table*}[htbp] 
\caption{ Results of Fitting Multi-wavelength SEDs in Figure \ref{fig:G}. A LogParabola model is used as electron injected spectrum which is defined as
dN/dE = N${_0}$(E/E${_0}$)$^{(-\alpha-\beta*log(E/E{_0}))}$, where E${_0}$ is chosen as 90 MeV.}
\centering
\begin{tabular}{c c c c c}
\hline
 Activity &Parameters & Symbol & Values& Activity period (days) \\
\hline
 & Min Lorentz factor of injected electrons & $\gamma_{min}$ & 3.5  \\
 & Max Lorentz factor of injected electrons & $\gamma_{max}$ & 7.5$\times$10$^{3}$  \\
 & BLR temperature & $T'_{blr}$ & 5$\times$10$^{4}$ K  \\
 & BLR photon density & $U'_{blr}$ & 1 erg/cm$^{3}$ \\
 & Disk temperature & $T'_{disk}$ & 2.6$\times$10$^{6}$ K  \\
 & Disk photon density & $U'_{disk}$ & 3.7$\times$10$^{-7}$ erg/cm$^{3}$ \\
 & Size of the emission region & R & 6.5$\times$ 10$^{16}$ cm   \\
 & Doppler factor of emission region& $\delta$ & 35 \\
 & Lorentz factor of emission region& $\Gamma$ & 15 \\
\hline
\hline
 Pre-flare&&&  \\
 & Spectral index of injected electron spectrum (LP) & $\alpha$ & 1.9 \\
 & Curvature parameter of LP electron spectrum & $\beta$ & 0.08 \\
 & magnetic field in emission region & B & 4.0 G & 56  \\
 & luminosity in injected electrons & $L_{e}$ & 1.78$\times$ 10$^{42}$ erg/sec  \\
 \hline 
 Rising &&& \\
 & Spectral index of injected electron spectrum (LP) & $\alpha$ & 1.8 \\
 & Curvature parameter of LP electron spectrum & $\beta$ & 0.08 \\
 & magnetic field in emission region & B & 4.1 G & 29  \\
 & luminosity in injected electrons & $L_{e}$ & 6.94$\times$ 10$^{42}$ erg/sec  \\
 \hline 
 Decaying &&& \\
 & Spectral index of injected electron spectrum (LP) & $\alpha$ & 1.8 \\
 & Curvature parameter of LP electron spectrum & $\beta$ & 0.08 \\
 & magnetic field in emission region & B & 4.1 G & 22 \\
 & luminosity in injected electrons & $L_{e}$ & 8.76$\times$ 10$^{42}$ erg/sec  \\
 \hline
  Flare-1 &&&  \\
 & Spectral index of injected electron spectrum (LP) & $\alpha$ & 1.7 \\
 & Curvature parameter of LP electron spectrum & $\beta$ & 0.02 \\ 
 & magnetic field in emission region & B & 4.2 G & 5  \\
 & luminosity in injected electrons & $L_{e}$ & 1.27$\times$ 10$^{44}$ erg/sec  \\
 \hline 
 Flare-2 &&& \\
 & Spectral index of injected electron spectrum (LP) & $\alpha$ & 1.7 \\
 & Curvature parameter of LP electron spectrum & $\beta$ & 0.02 \\
 & magnetic field in emission region & B & 4.1 G  & 8 \\
 & luminosity in injected electrons & $L_{e}$ & 5.0$\times$ 10$^{43}$ erg/sec  \\
 \hline 
 Flare-3 &&& \\
 & Spectral index of injected electron spectrum (LP) & $\alpha$ & 1.7 \\
 & Curvature parameter of LP electron spectrum & $\beta$ & 0.02 \\
 & magnetic field in emission region & B & 4.2 G & 8 \\
 & luminosity in injected electrons & $L_{e}$ & 9.04$\times$ 10$^{43}$ erg/sec  \\
 \hline
 Flare-4 &&& \\
 & Spectral index of injected electron spectrum (LP) & $\alpha$ & 1.7 \\
 & Curvature parameter of LP electron spectrum & $\beta$ & 0.02 \\
 & magnetic field in emission region & B & 4.2 G & 7 \\
 & luminosity in injected electrons & $L_{e}$ & 8.91$\times$ 10$^{43}$ erg/sec  \\
\hline
\hline
 \end{tabular}
\label{Table:T4}
\end{table*} 

\begin{figure*}
\centering 
\subfigure[]{%
\label{fig:ex3-a}%
\includegraphics[scale=0.40]{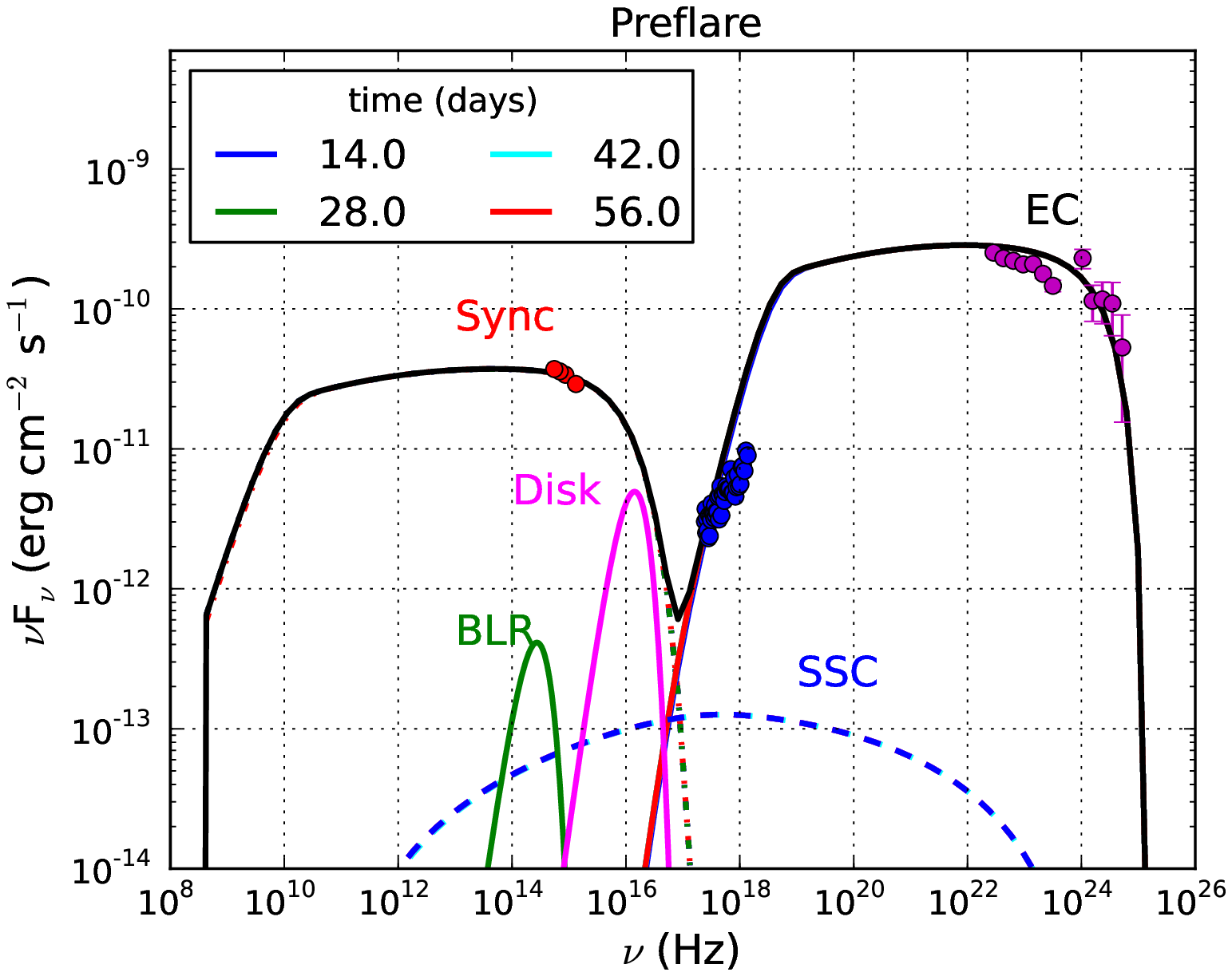}}%
\subfigure[]{%
\label{fig:ex3-b}%
\includegraphics[scale=0.40]{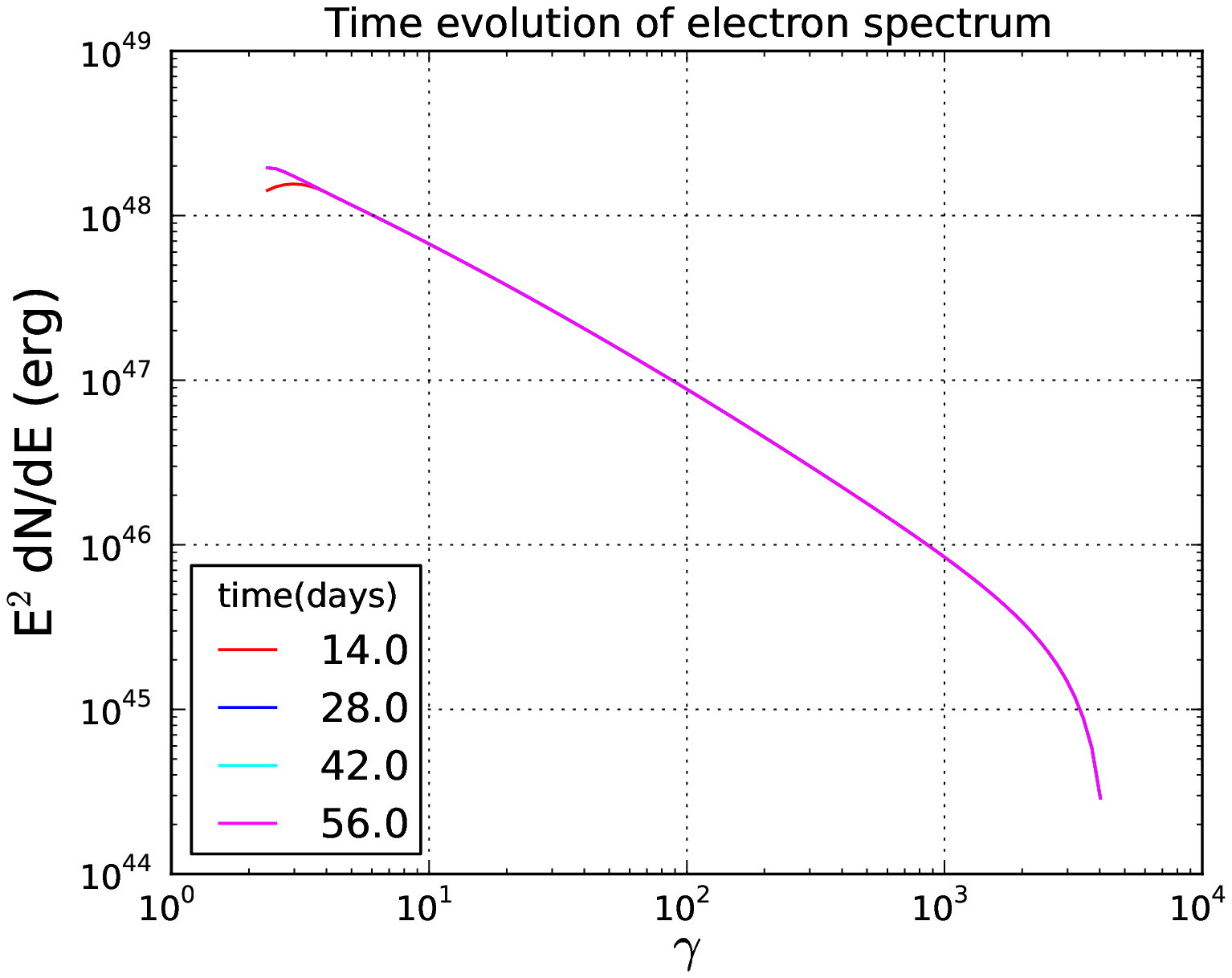}}%
\qquad 
\centering
\subfigure[]{%
\label{fig:ex3-c}%
\includegraphics[scale=0.40]{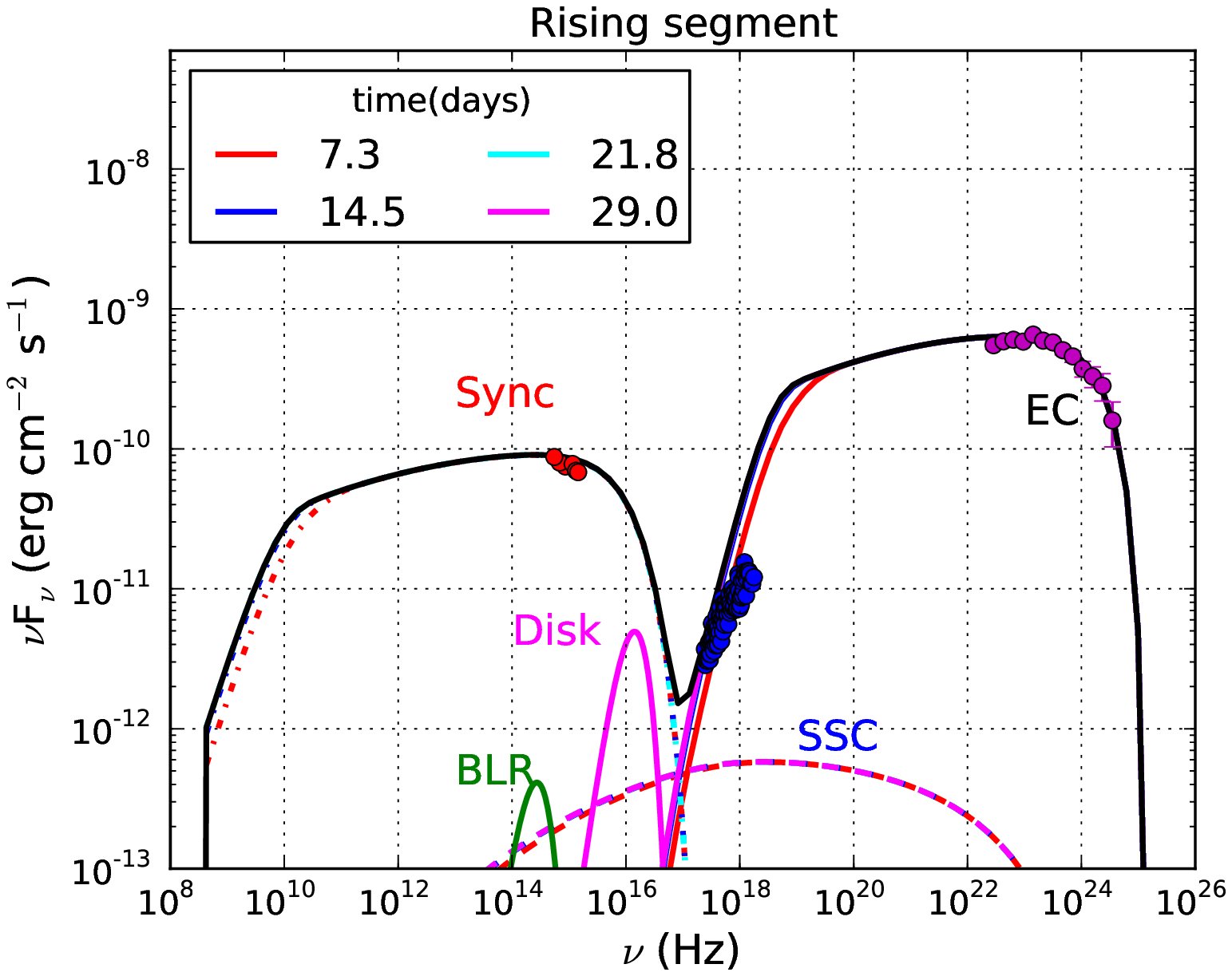}}%
\subfigure[]{%
\label{fig:ex3-d}%
\includegraphics[scale=0.40]{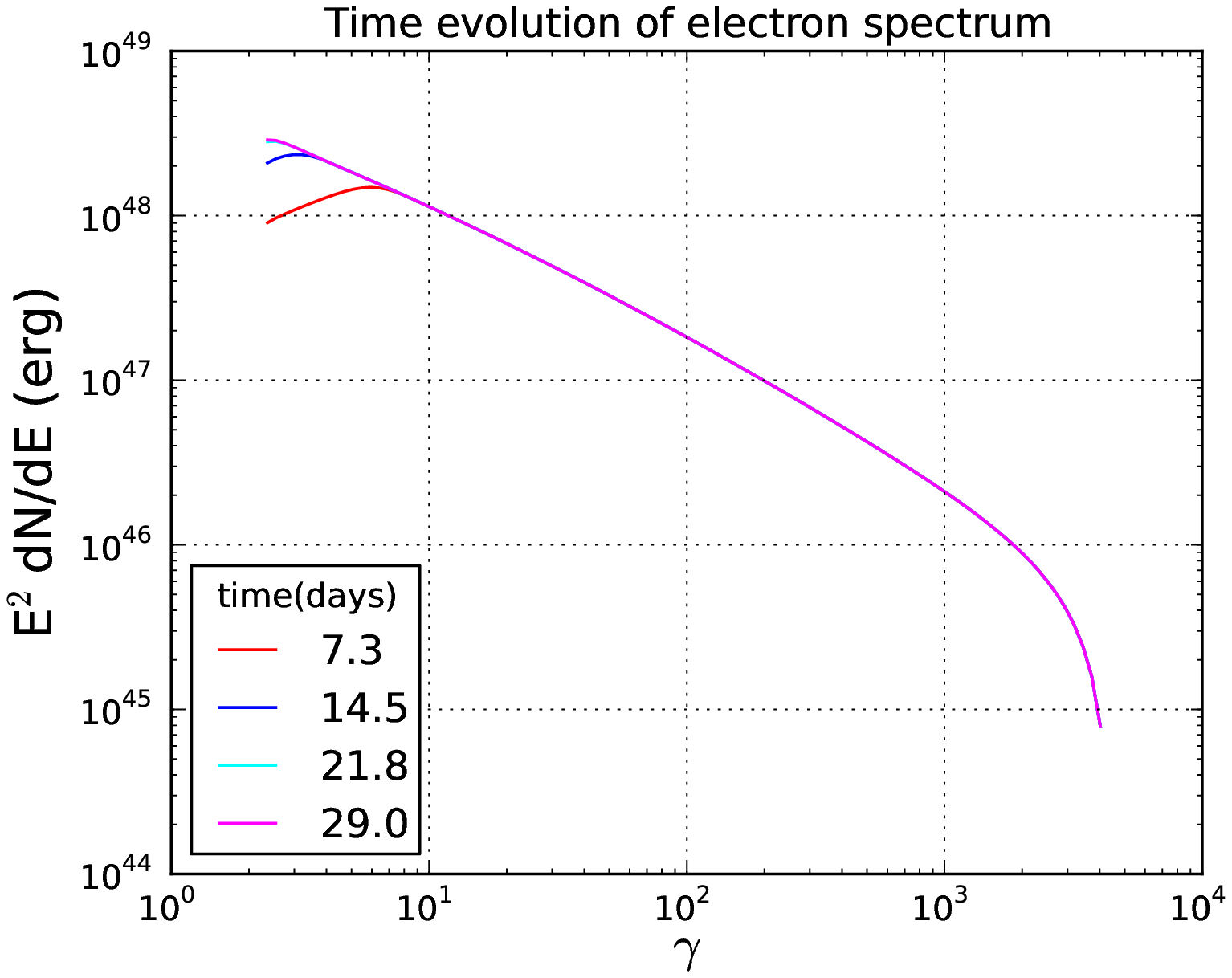}}%
\qquad 
\centering
\subfigure[]{%
\label{fig:ex3-e}%
\includegraphics[scale=0.40]{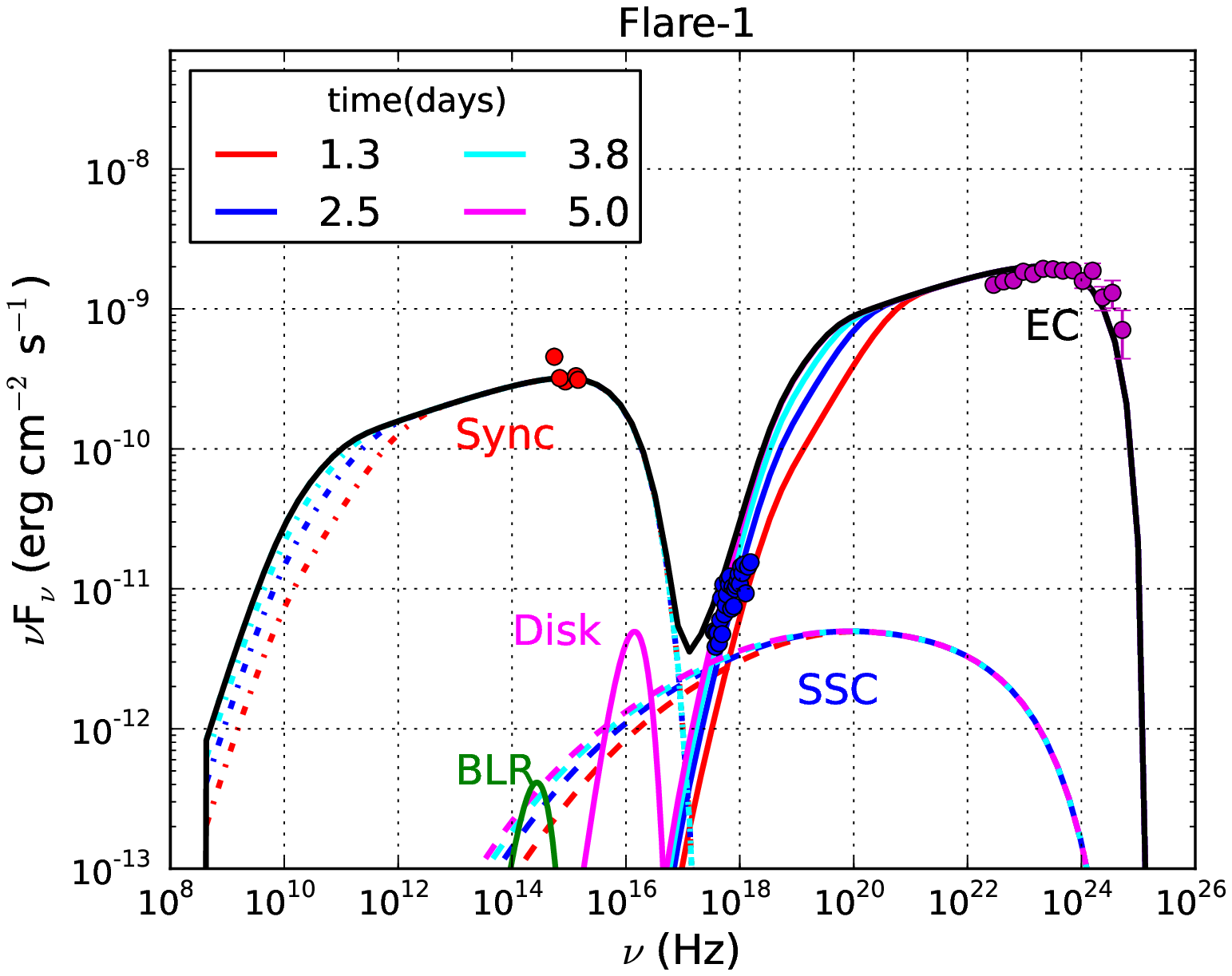}}%
\subfigure[]{%
\label{fig:ex3-f}%
\includegraphics[scale=0.40]{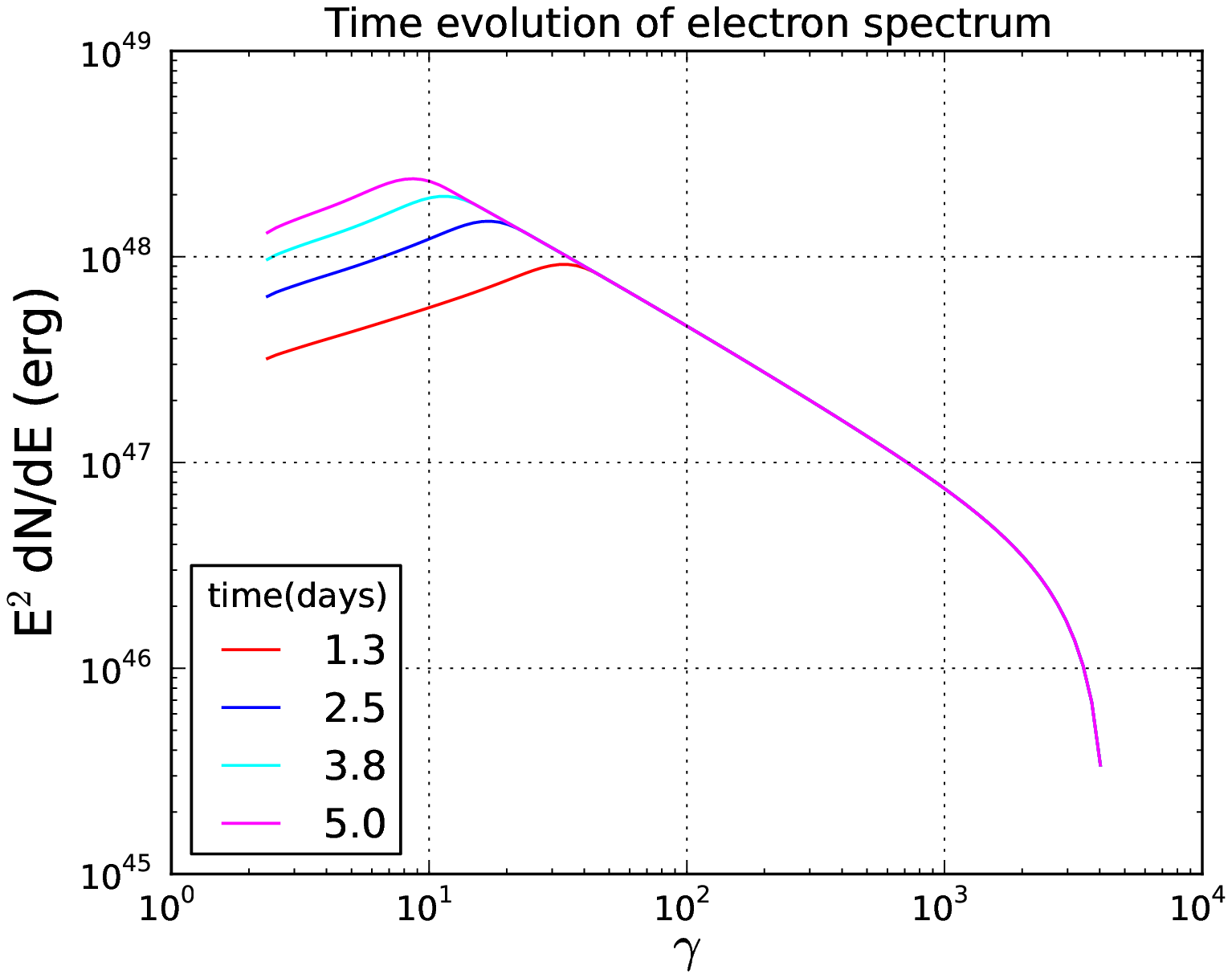}}%
\qquad 
\centering 
\subfigure[]{%
\label{fig:ex3-g}%
\includegraphics[scale=0.40]{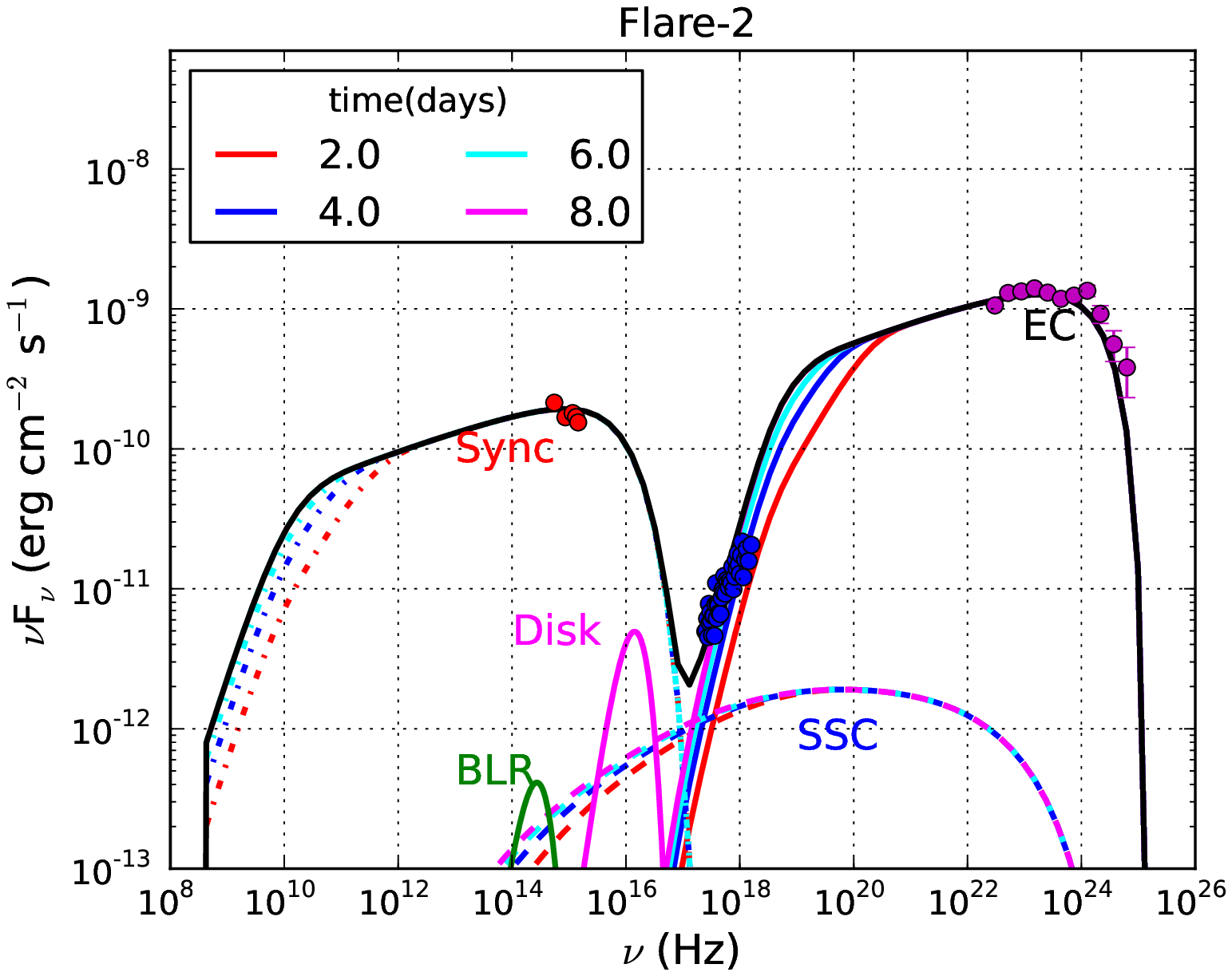}}%
\subfigure[]{%
\label{fig:ex3-h}%
\includegraphics[scale=0.40]{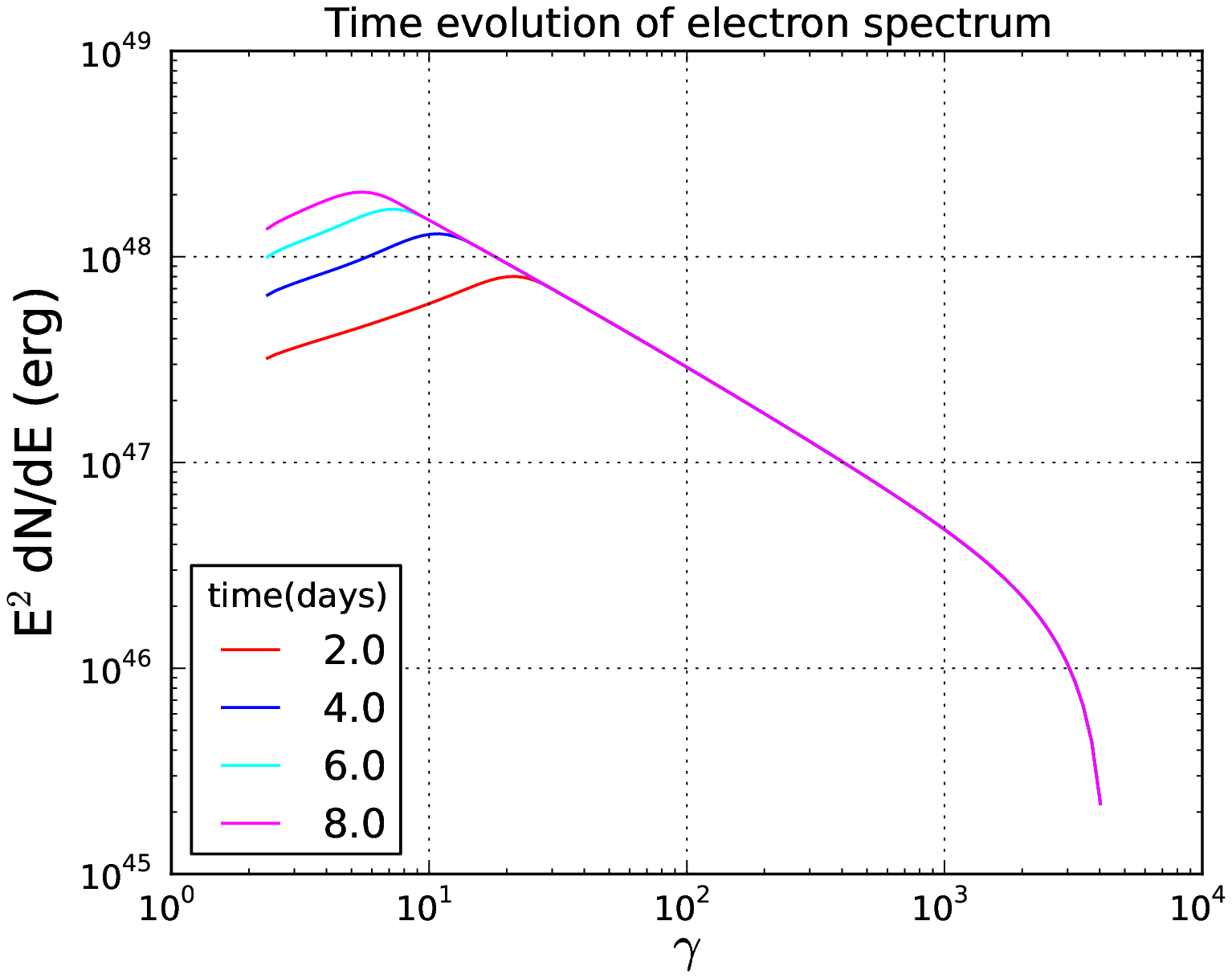}}%
\end{figure*}
\clearpage 
\bibliographystyle{plain}

\end{document}